\documentclass[12pt, draftclsnofoot, onecolumn]{IEEEtran}  %default 1.5 apace
\usepackage{cite}
\usepackage{amsmath,amssymb,amsfonts,,lipsum}
\usepackage{algorithmic}
\usepackage{graphicx}
\usepackage{subfigure}
\usepackage{textcomp}
\usepackage{epstopdf}
\usepackage{xcolor}
\usepackage{caption}
\usepackage{array}
\usepackage{subeqnarray}
\usepackage{stfloats}
\usepackage{CJK}
\usepackage{changepage}
\usepackage{ulem}

\usepackage{enumerate}

\usepackage[ruled,vlined]{algorithm2e}
\newtheorem{theorem}{Theorem}

\newtheorem{property}{Property}
\newtheorem{corollary}{Corollary}

\newtheorem{remark}{Remark}
\newtheorem{definition}{Definition}

\newtheorem{case1}{Case}
\newtheorem{case2}{Case}
\def\BibTeX{{\rm B\kern-.05em{\sc i\kern-.025em b}\kern-.08em
   T\kern-.1667em\lower.7ex\hbox{E}\kern-.125emX}}
\let\sss= \scriptscriptstyle

\ifCLASSINFOpdf
\else
\fi
\begin{document}
\captionsetup[figure]{name={Fig.},labelsep=period}
\normalem
\title{RIS-Aided Wireless Communications: \\Extra Degrees of Freedom via \\Rotation and Location Optimization}

\author
    {
        Yajun Cheng, Wei Peng, \IEEEmembership{Senior Member, IEEE}, Chongwen Huang, \IEEEmembership{Member, IEEE}, George C. Alexandropoulos, \IEEEmembership{Senior Member, IEEE}, Chau Yuen,~\IEEEmembership{Fellow, IEEE}, and M$\acute{\text{e}}$rouane Debbah, \IEEEmembership{Fellow, IEEE}% <-this % stops a space
        \thanks{Yajun Cheng is with the School of Electronic Information and Communications, Huazhong University of Science and Technology, and Wei Peng is with the School of Cyber Science and Engineering,  Huazhong University of Science and Technology, respectively. And they are also with the Research Center of 6G Mobile Communications, Huazhong University of Science and Technology, 1037 Luoyu Road, Wuhan, 430074, China. (e-mail: yajuncheng@hust.edu.cn, pengwei@hust.edu.cn).}
        \thanks{Chongwen Huang is with College of Information Science and Electronic Engineering, Zhejiang University, Hangzhou 310027, China, and with International Joint Innovation Center, Zhejiang University, Haining 314400, China, and also with Zhejiang Provincial Key Laboratory of Info. Proc., Commun. \& Netw. (IPCAN), Hangzhou 310027, China. (e-mail: chongwenhuang@zju.edu.cn).}
        \thanks{George C. Alexandropoulos is with the Department of Informatics and Telecommunications, National and Kapodistrian University of Athens, 15784 Athens, Greece. (e-mail: alexandg@di.uoa.gr).}
        \thanks{Chau Yuen is with the Engineering Product Development (EPD) Pillar, Singapore University of Technology and Design, Singapore 487372 (e-mail: yuenchau@sutd.edu.sg).}
        \thanks{M$\acute{\text{e}}$rouane Debbah is with the Technology Innovation Institute, 9639 Masdar City, Abu Dhabi, United Arab Emirates and with the CentraleSup$\acute{\text{e}}$lec, University Paris-Saclay, 91192 Gif-sur-Yvette, France. (e-mail: merouane.debbah@tii.ae).}
%        \thanks{The corresponding author of this work is Wei Peng.}
        % <-this % stops a space
    }
\maketitle

    \begin{abstract}
      We consider the extra degree of freedom offered by the rotation of the reconfigurable intelligent surface (RIS) plane and investigate its potential in improving the performance of RIS-assisted wireless communication systems. By considering radiation pattern modeling at all involved nodes, we first derive the composite channel gain and present a closed-form upper bound for the system ergodic capacity over cascade Rician fading channels. Then, we reconstruct the composite channel gain by taking the rotations at the RIS plane, transmit antenna, and receive antenna into account, and extract the optimal rotation angles after investigating their impacts on the capacity. Moreover, we present a location-dependent expression of the ergodic capacity and investigate the RIS deployment strategy, i.e. the joint rotation adjustment and location selection. Finally, simulation results verify the accuracy of the theoretical analyses and deployment strategy. Although the RIS location has a big impact on the performance, our results showcase that the RIS rotation plays a more important role. In other words, we can obtain a considerable improvement by properly rotating the RIS rather than moving it over a wide area. For instance, we can achieve more than 200\% performance improvement through rotating the RIS by 42.14$^{\circ}$, while an 150\% improvement is obtained by shifting the RIS over 400 meters.
    \end{abstract}


% \fi
\IEEEpeerreviewmaketitle

    \section{Introduction}

     \IEEEPARstart{W}{ith} the inherent ability in tailoring the radio propagation environment, reconfigurable intelligent surfaces (RIS) technology has overwhelmingly emerged as one of the most promising approaches for future wireless communications \cite{MRenzoSREbyAIRIS,MAElMossallamyRISPrinciples,ECalvaneseRIS6Gsmart}. The RIS is a two-dimensional surface composed of a large number of passive units arranged mostly in sub-wavelength inter-element spacing with the property of manipulating the electromagnetic waves, such as scattering, reflection, and absorption \cite{MRenzoRISSurvey,CHuangHolographicRIS,GCAlexandropoulosRISpotential}.

     Currently, a plethora of studies have concentrated on the discrete phase shift designs \cite{YChengPhaseAdjustmentIRS,SAbeywickramaRISPracticaPhase}, joint active and passive beamforming optimizations \cite{CHuangRISDeepReinforcement,BDiRISHybridBeamforming}, and performance analyses, such as signal-to-noise ratio (SNR) coverage probability \cite{ZCuiSNRAnalysisRISCSRayleigh,LYangCoverProbSNRRISRayleigh}, energy/spectral efficiency \cite{CHuangRISEE,SZhouSEEEIRSMISOCS}, and capacity characterization\cite{AAABoulogeorgoscapacityRISRayleigh,DLiCapacityRISCSPhaseErrorsRayleigh,YHanRISStatisticalCSI,SZhangCapacityRISMIMOLOS,QTaoRISPerformanceSISORician,AMSalhabPerformanceRISRician,DSelimisRISNakagami,XGanUserSelectionRISCSRician,NSPeroviRISAchievablerate,ALMoustakascapacityRIS,COuyangergodiccapacityRISMIMO}. In the capacity analysis context, \cite{AAABoulogeorgoscapacityRISRayleigh} presented an approximation of the ergodic capacity of RIS-aided communication system, whilst \cite{DLiCapacityRISCSPhaseErrorsRayleigh} discussed the capacity with phase errors. However, these conclusions are based on the Rayleigh fading channels. To cover the scenario with line-of-sight (LOS) links, some researchers paid attention to the capacity analyses over Rician fading channels \cite{YHanRISStatisticalCSI,XGanUserSelectionRISCSRician,SZhangCapacityRISMIMOLOS,COuyangergodiccapacityRISMIMO,NSPeroviRISAchievablerate,QTaoRISPerformanceSISORician,ALMoustakascapacityRIS,AMSalhabPerformanceRISRician}. For instance, the authors studied the capacity of RIS-aided MISO system of single-user and multi-user scenarios by exploiting statistical CSI in \cite{YHanRISStatisticalCSI} and\cite{XGanUserSelectionRISCSRician}, respectively. The works in \cite{SZhangCapacityRISMIMOLOS} and \cite{NSPeroviRISAchievablerate} explored the achievable rate of RIS-aided MIMO systems by jointly optimizing the transmitted signal covariance matrix and RIS phase via an alternating optimization algorithm and an iterative projected gradient algorithm, respectively. The authors in \cite{COuyangergodiccapacityRISMIMO} provided the upper and lower bounds of ergodic capacity of a MIMO system based on random matrix theory, as well as the asymptotic analyses, while the authors in \cite{QTaoRISPerformanceSISORician} and \cite{AMSalhabPerformanceRISRician} carried out the upper bound analysis on the ergodic or average capacity using the Laguerre polynomial. In \cite{ALMoustakascapacityRIS}, an asymptotic closed-form expression, in the large-antenna limit, for the mutual information of a multi-antenna transmitter-receiver pair in the presence of multiple RISs was presented using the random matrix and replica theories.
     In addition, the works in \cite{DSelimisRISNakagami} studied the capacity of RIS-aided MISO system over Nakagami-$m$ fading channels.

     Most of the existing analyses are based on the premise that RIS is deployed at a given location. According to \cite{MAElMossallamyRISPrinciples} and \cite{QWuRIStutorial}, the sum-distance or product-distance based path-loss may have a significant impact on RIS-aided communication performances.
     Hence, some researchers have paid attention to the deployment issues \cite{MAKishkRISRandblockDeployment,BZhengDoubleRISMIMO,YGaoDistrRISMISO,ZKangRISoptdeploy,SZhangRISCapacityDeployment}. For example, it was found in \cite{MAKishkRISRandblockDeployment} that the large-scale deployment of the RISs is capable of improving the coverage regions. The work in \cite{BZhengDoubleRISMIMO} showed the superiority of the double-RIS deployment than a single-RIS. In addition, the authors in \cite{YGaoDistrRISMISO} showcased that uniformly distributed deployment of RISs outperforms the centralized deployment with the same scale. Moreover,\cite{ZKangRISoptdeploy} studied the RIS deployment strategy of a RIS-aided relay system, including single-RIS and multi-RIS deployments, and showed that the multi-RIS deployment can obtain a higher system capacity. However, the study in \cite{SZhangRISCapacityDeployment} revealed that the centralized deployment is superior to the distributed one by characterizing the capacity. These works focused on the comparisons between centralized (single-RIS) and distributed (multiple-RIS) deployments, rather than location optimization. Although, the study in \cite{QWuRIStutorial} showed that the closer the RIS is to the user or base station, the larger the SNR can be yielded, the impact of RIS orientation is neglected because the RIS unit was seen as an isotropic radiator.

     Additionally, there are some studies that involve the discussion of RIS orientation. For instance, the study in \cite{ZCuiSNRAnalysisRISCSRayleigh} showed that the effective area of RIS has a significant impact on the SNR, where the effective area is sensitive to its orientation, but the optimal orientation was not revealed. The authors in \cite{SZengRISOrientaLocation} indicated that the RIS should be deployed vertically to the direction from the base station to RIS for coverage extension, while the authors in \cite{XChengJointoptRIS} suggested that the RIS ought to be optimally oriented to make it in specular reflection state to maximize the received power. However, these analyses are based on the assumption of LOS channel models, which is not universal and is not applicable in scenarios where non-LOS (NLOS) links exist. Additionally, \cite{WTangRISPathlossmeasurement} proposed a RIS-aided path loss model and validated its correctness via simulations and trial measurements. Nevertheless, it was pointed out that there is about 3 dB difference between the measurement and theoretical results. This gap might partly result from the assumption that both the incident and reflecting directions are perpendicular to the RIS.

     In summary, the prior literature overlooked some crucial inherent characteristics of RIS, such as the directivity of the RIS, which will influence the effective radiation area and lead to different conclusions. Intuitively, the effective area of RIS that is perpendicular to the direction of propagation varies with its rotation and location. Hence, the isotropic radiation hypothesis of RIS might yield over-optimistic performance. On the other hand, although RIS orientation has been discussed in some existing works, the conclusions are not universal and the applicable scenarios are limited. Moreover, both the orientation and location of RIS need to be considered simultaneously, as both have a significant impact on the performance.

     Against the above background, in this paper, we dedicate our efforts to derive a general expression of the ergodic capacity for RIS-aided communication systems, where both the LOS and NLOS links are considered, and provide a new degree of freedom in optimizing RIS-aided wireless channel by taking the radiative characteristics of RIS into account. Based on the channel model, we explore the deployment strategy of RIS, including the optimizations of RIS rotation and location. The main contributions are summarized as follows:
        \begin{enumerate}
            \item[$\bullet$] We introduce a practical channel expression by considering the radiative characteristics of RIS, which provides a new degree of freedom in optimizing the RIS-aided wireless communications.

            \item[$\bullet$] We establish a general model of RIS-aided point-to-point wireless communication system with directional transmit antenna (Tx-Ant) and receive antenna (Rx-Ant). This system model can be applied to unmanned aerial vehicle (UAV), satellite, and  terahertz communication systems.

            \item[$\bullet$] We provide a tight upper bound of the ergodic capacity over cascade Rician fading channels. The expression of the capacity can be exactly characterized by Hypergeometric function, and covers the cases of double-LOS and double-Rayleigh links, which can be used for the RIS of any scale.

            \item[$\bullet$] We excavate the impact of the Tx-Ant, Rx-Ant and RIS rotations on the performance and investigate their optimal rotation angles at any location of the RIS. It is crucial to place the Tx-Ant and Rx-Ant by aligning their radiation peaks to the RIS centre. And we show that improper rotation of RIS will lead to significant performance degradation.

            \item[$\bullet$] We present an expression of the ergodic capacity that is a function of the RIS location and propose two algorithms for the RIS deployment, including the optimal location and effective region of the RIS deployment that ensures the SNR is not less than a given threshold.
        \end{enumerate}

     The rest of the paper is organized as follows. The system and channel models are described in Section II. The downlink ergodic capacity over cascade Rician channels is investigated in Section III. Then, the optimization of rotation angles and RIS location are explored in Section IV. The simulation results and conclusions are provided in Sections V and VI, respectively.

     Notation: In this paper, $\mathcal{CN}(0,\sigma^{2})$ represents the Complex Gaussian distribution with zero mean and variance $\sigma^{2}$. $\mathbb{E}\left\{\cdot\right\}$ means the statistical expectation of a random variable and $\left|\cdot\right|$ denotes the modulus operator. $\log\left(\cdot\right)$ and $e$ are the logarithm operator and natural constant, respectively. In addition, ${_{p}F_{q}}\left(a_{1},\cdots,a_{p};b_{1},\cdots,b_{q};z\right)$ represents the generalized Hypergeometric function \footnote{The Hypergeometric function is defined by the power series as ${_{p}F_{q}}\left(a_{1}\dots a_{p};b_{1}\dots b_{q};z\right)=\sum_{k=0}^{\infty}\frac{\left(a_{1}\right)_{k}\dots \left(a_{p}\right)_{k}}{\left(b_{1}\right)_{k}\dots \left(b_{q}\right)_{k}}\frac{z^{k}}{k!}$, with $\left(a_{p}\right)_{k}=a_{p}\left(a_{p}+1\right)\left(a_{p}+k-1\right)$ \cite[Eq. (16.2.1)]{FWOlverNISTHandbook}.} and $I_{\nu}(\cdot)$ stands for the $\nu^{th}$ order modified Bessel functions of the first kind. Moreover, $\chi_{K}^{2}\left(\lambda_{0}\right)$ represents the non-central chi-squared distribution with $K$ degrees freedom and non-centrality parameter $\lambda_{0}$.
     \begin{table}[!h]
	\centering
	\caption{List of variables.}
    \renewcommand\arraystretch{0.8}
  	\label{tab:notations}
	%\begin{tabularx}{0.8\textwidth}{c|X}
    \begin{tabular}{m{1.3cm}<{\centering}|m{6cm}||m{1.4cm}<{\centering}|m{5.5cm}}
		\hline
		\hline
		\bf{Variable} & \bf{Description}                                                                      & \bf{Variable} & \bf{Description}\\
        \hline
        $N$ &  The number of the RIS units                                                                    & $P_{t}$ &  The transmit power at Tx-Ant \\
       	\hline
		$g_{n}$ &  The channel between Tx-Ant and $n^{th}$ RIS unit                                             & $\widehat{g}_{n}$/$\widetilde{g}_{n}$ &  The LOS/NLOS component of $g_{n}$\\
		\hline
		$z_{n}$ & The channel between Rx-Ant and $n^{th}$ RIS unit                                              & $\overline{z}_{n}$/$\widetilde{z}_{n}$ &  The LOS/NLOS component of $z_{n}$\\
		\hline
		$\widehat{g}_{n}$ & The equivalent channel via $n^{th}$ RIS unit                                        & $\mathcal{N}_{0}$ &  The noise power at Rx-Ant\\
        \hline
		$K_{1}$ & The Rician factor of the channel between Tx-Ant and RIS                                     & $K_{2}$ &  The Rician factor of the channel between Rx-Ant and RIS\\
		\hline
        $C/\overline{C}$ &  The capacity/capacity upper bound of the proposed system                          & $\mathcal{T}_{n}$ & The reflection coefficient introduced by $n^{th}$ RIS unit \\
        \hline
        $\overline{C}_{\text{R}}$ &  The $\overline{C}$ with rotation at RIS                                  & $\mathcal{SNR}$ &  The signal-to-noise ratio at Rx-Ant\\
		\hline
        $F_{t}\left(\theta,\varphi\right)$ & The normalized field pattern of Tx-Ant                           & $U_{t}\left(\theta,\varphi\right)$ & The normalized power pattern of Tx-Ant\\
        \hline
        $F_{r}\left(\theta,\varphi\right)$ & The normalized field pattern of Rx-Ant                           & $U_{r}\left(\theta,\varphi\right)$ & The normalized power pattern of Rx-Ant\\
        \hline
        $F_{u}\left(\theta,\varphi\right)$ & The normalized field pattern of RIS unit                         & $U_{u}\left(\theta,\varphi\right)$ & The normalized power pattern of RIS unit\\
        \hline
        $q_{t}/q_{r}/q_{u}$ & The directivity parameters of Tx-Ant/Rx-Ant/RIS unit                            & $D_{t}/D_{r}/D_{u}$ & The maximum directivity of Tx-Ant/Rx-Ant/RIS unit\\
		\hline
        $\rho_{0}$  & The path-loss factor at $d=1 m$                                                         & $\rho_{cc}$ &  The  composite channel gain\\
		\hline
        $\rho_{t,n}/\rho_{t,c}$ & The path gain between Tx-Ant and $n^{th}$ RIS unit/RIS center                 & $d_{t,n}/d_{t,c}$ & The distance between Tx-Ant and $n^{th}$ RIS unit/RIS center\\
		\hline
        $\rho_{n,r}/\rho_{c,r}$ & The path gain between $n^{th}$ RIS unit/RIS center and Rx-Ant                 & $d_{n,r}/d_{c,r}$ & The distance between $n^{th}$ RIS unit/RIS center and Rx-Ant\\
		\hline
		$\theta_{t\text{0}}/\theta_{r\text{0}}/\theta_{\text{0}}$ & The rotation angle at the Tx-Ant/Rx-Ant/RIS  & $h_{t}$/$h_{r}$ & The height of the Tx-Ant/Rx-Ant\\
        \hline
		$R$ & The horizontal distance between the Tx-Ant and Rx-Ant                                           & $r$ & The vertical distance between the RIS and the XOZ plane\\
        \hline
        $l$ & The vertical distance between the RIS and the YOZ plane                                         & $h$ & The altitude of RIS (vertical distance to the XOY plane)\\
        \hline		
        \hline
	\end{tabular}
    \end{table}

    \section{System and Channel Models}
\begin{figure}[!t]
            \centering
            \subfigure[Communications assisted by a RIS.]{
            \label{Fig_1_1}
            \includegraphics[width=0.45\linewidth]{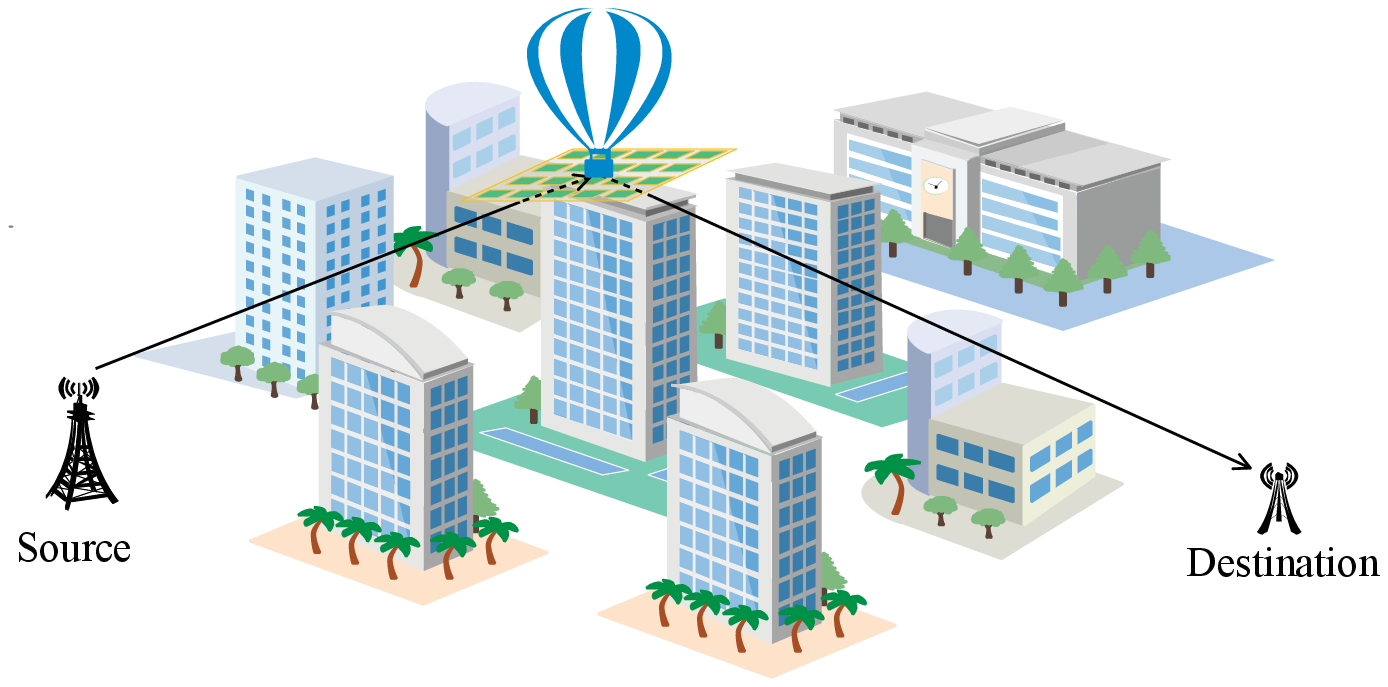}}
            \hspace{0.2in}
            \subfigure[The point-to-point RIS-aided communication system.]{
            \label{Fig_1_2}
            \includegraphics[width=0.45\linewidth]{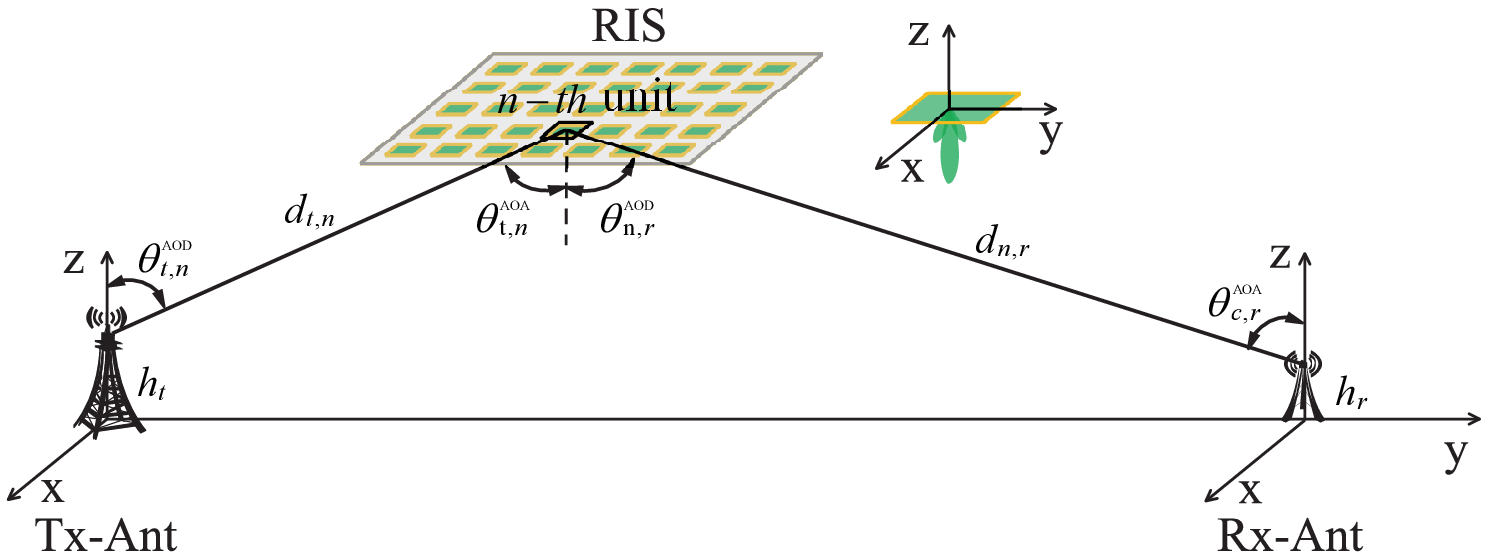}}
            \caption{The considered system model.}
            \label{Fig_1}
        \end{figure}

        In this paper, we consider a typical point-to-point RIS-aided wireless communication system. As depicted in Fig. \ref{Fig_1} \subref{Fig_1_1}, the RIS performs as a relay to connect the communication between the source node (such as a base station or an access point) and destination node (say a mobile equipment) because the direct link is blocked by the dense and tall buildings. The gain generated by multi-hop channels through building facades is too weak and is usually neglected. In light of the advantages of RIS in favor of communications, we propose a RIS-aided communication system, where the RIS is equipped on a hot air balloon or UAV and supposed to be deployed in somewhere to assist the communication. Please note that, in this paper, we assume that the RIS can be rotated by manipulating a mechanical mechanism underneath it, and for the case of hot air balloon or UAV, its location can be adjusted by moving the hot air balloon or UAV. Moreover, a directional single antenna with adjustable direction is considered at both the source and destination nodes without loss of generality. Furthermore, we assume that the locations and rotations of the Tx-Ant, Rx-Ant, and RIS are all known.

        The considered RIS-aided communication scenario can be simplified as illustrated in Fig. \ref{Fig_1} \subref{Fig_1_2}. The source and destination nodes can be seen as a transmitter-receiver (Tx-Rx) pair with directional single antenna, which can be rotated. For easy exposition, we assume that the Tx-Rx pair is located on the Y-axis in Cartesian coordinates with the height $h_{t}$ and $h_{r}$, respectively. The source node is located at the origin. Additionally, the Tx-Ant and Rx-Ant are both assumed to be initially deployed with their major lobes directed along the Z-axis. Moreover, the RIS consisting of $N$ directional reflection units is placed above the Tx-Rx pair at an altitude $h$. Please note that the RIS is assumed to be paralleled to the X-axis-Origin-Y-axis (XOY) plane initially. Furthermore, the direct link between the Tx-Ant and Rx-Ant is blocked and neglected.

            In practice, the RIS is usually deployed in places to optimize the communication channels by providing LOS links. Nevertheless, the effect of the NLOS links is not negligible, especially in scenarios full of scatterers, such as buildings and trees. Without loss of generality, the channel $g_{n}$ between the Tx-Ant and $n^{th}$ RIS unit, and the channel $z_{n}$ between $n^{th}$ RIS unit and the Rx-Ant are assumed as Rician fading channels, and they are respectively formulated as
            \begin{equation} \label{eq:channel-hn}
            \begin{aligned}
                 g_{n}=\sqrt{\rho_{t,n}}\left(\sqrt{\frac{K_{1}}{1+K_{1}}}\overline{g}_{n}+\sqrt{\frac{1}{1+K_{1}}}\widetilde{g}_{n}\right),
            \end{aligned}
            \end{equation}
            \begin{equation} \label{eq:channel-zn}
            \begin{aligned}
                 z_{n}=\sqrt{\rho_{n,r}}\left(\sqrt{\frac{K_{2}}{1+K_{2}}}\overline{z}_{n}+\sqrt{\frac{1}{1+K_{2}}}\widetilde{z}_{n}\right),
            \end{aligned}
            \end{equation}
        where $\overline{g}_{n}=e^{-j\frac{2\pi}{\lambda}d_{t,n}}$ and $\overline{z}_{n}=e^{-j\frac{2\pi}{\lambda}d_{n,r}}$ denote the LOS links with $d_{t,n}$ representing the distance between the Tx-Ant and $n^{th}$ RIS unit, and $d_{n,r}$ being the distance between $n^{th}$ RIS unit and the Rx-Ant. Moreover, $\widetilde{g}_{n}\sim\mathcal{CN}\left(0,1\right)$ and $\widetilde{z}_{n}\sim \mathcal{CN}\left(0,1\right)$ denote the scattered components indicating the NLOS links. Furthermore, $K_{1}$ and $K_{2}$ are the Rician factors. By varying $K_{1}$ and $K_{2}$, $h_{n}$ and $z_{n}$ capture a typical wireless communication channel spanning from a Rayleigh fading channel ($K_{1}=0$ or $K_{2}=0$) to a LOS link ($K_{1}\rightarrow\infty$ or $K_{2}\rightarrow\infty$). Finally, $\rho_{t,n}$ and $\rho_{n,r}$ denote the large scale path loss.

        As for the large scale fading, the typical distance-dependent propagation model is adopted, i.e. $\rho\left(d\right)=\rho_{0}/d^{\upsilon}$, where $\rho_{0}$ represents the path-loss factor at the reference distance $d_{0}=1$ meter and $\upsilon$ denotes the path-loss exponent. Although the direct link between the Tx-Ant and Rx-Ant is blocked in the dense scenario, the RIS is to be assumed deployed in the air to facilitate communications by providing additional links or connections, which contains the LOS components. For the ground-to-air scenario, the path-loss exponent $\upsilon$ is typically assumed as 2 \cite{HLuAerialRIS,MHuaUAVAidedWSN,XCaoRISAerialTaskLearning}. Therefore, the path gain between the Tx-Ant and $n^{th}$ RIS unit is given by
            \begin{equation} \label{eq:pathgain-BS-RISn}
            \begin{aligned}
                 \rho_{t,n}=\frac{\rho_{0}G_{t}\left(\theta_{t,n}^{\sss{\text{AOD}}},\varphi_{t,n}^{\sss{\text{AOD}}}\right)
                 G_{u}\left(\theta_{t,n}^{\sss{\text{AOA}}},\varphi_{t,n}^{\sss{\text{AOA}}}\right)}{d_{t,n}^{2}},
            \end{aligned}
            \end{equation}
        where $\theta_{t,n}^{\sss{\text{AOD}}}$ and $\varphi_{t,n}^{\sss{\text{AOD}}}$ represent the azimuth and elevation angles of departure (AODs) from the Tx-Ant to $n^{th}$ RIS unit. Analogously, $\theta_{t,n}^{\sss{\text{AOA}}}$ and $\varphi_{t,n}^{\sss{\text{AOA}}}$ represent the elevation and azimuth angles of arrival (AOAs) from the Tx-Ant to $n^{th}$ RIS unit. Moreover,
        $G_{t}\left(\theta_{t,n}^{\sss{\text{AOD}}},\varphi_{t,n}^{\sss{\text{AOD}}}\right)$ is the radiation gain of the Tx-Ant in the direction of $\left(\theta_{t,n}^{\sss{\text{AOD}}},\varphi_{t,n}^{\sss{\text{AOD}}}\right)$, while $G_{u}\left(\theta_{t,n}^{\sss{\text{AOA}}},\varphi_{t,n}^{\sss{\text{AOA}}}\right)$ denotes the radiation gain of the RIS unit in the direction of $\left(\theta_{t,n}^{\sss{\text{AOA}}},\varphi_{t,n}^{\sss{\text{AOA}}}\right)$.

        According to \cite{StutzmanAntennatheorydesign}, the radiation gain can be expressed as a function of angle by including the power pattern as $G\left(\theta, \phi\right)=e_{r}D\left|F\left(\theta,\phi\right)\right|^{2}$ \cite[Eq.(2-155)]{StutzmanAntennatheorydesign}. Here, $e_{r}$ is the radiation efficiency, which is close to 1, i.e. $e_{r}\approx 1$; $D$ is the maximum directivity and $F(\theta, \varphi)$ is the normalized field pattern \cite[Eq.(2-145)]{StutzmanAntennatheorydesign}.

        Naturally, the path gain between the Tx-Ant and $n^{th}$ RIS unit in (\ref{eq:pathgain-BS-RISn}) can be recast as
            \begin{equation} \label{eq:channel-gain-BS-RISn}
            \begin{aligned}
                 \rho_{t,n}&=\frac{\rho_{0}}{ d_{t,n}^{2}}D_{t}D_{u}\left|F_{t}\left(\theta_{t,n}^{\sss{\text{AOD}}},\varphi_{t,n}^{\sss{\text{AOD}}}\right)\right|^{2}\left|F_{u}\left(\theta_{t,n}^{\sss{\text{AOA}}},\varphi_{t,n}^{\sss{\text{AOA}}}\right)\right|^{2},
            \end{aligned}
            \end{equation}
       where $F_{t}\left(\theta,\varphi\right)$ and $F_{u}\left(\theta,\varphi\right)$ represent the normalized radiation pattern of the Tx-Ant and RIS unit, respectively. Moreover, $D_{t}$ and $D_{u}$ are the maximum directivity of the Tx-Ant and RIS unit, respectively.

        Analogously, the path gain between $n^{th}$ RIS unit and the Rx-Ant is given by
            \begin{equation} \label{eq:channel-gain-RISn-Rx}
            \begin{aligned}
                 \rho_{n,r}&=\frac{\rho_{0}}{ d_{n,r}^{2}}D_{u}D_{r}
                 \left|F_{u}\left(\theta_{n,r}^{\sss{\text{AOD}}},\varphi_{n,r}^{\sss{\text{AOD}}}\right)\right|^{2}\left|F_{r}\left(\theta_{n,r}^{\sss{\text{AOA}}},\varphi_{n,r}^{\sss{\text{AOA}}}\right)\right|^{2},
            \end{aligned}
            \end{equation}
        where $D_{r}$ and $F_{r}\left(\theta,\varphi\right)$ represent the maximum directivity and the normalized radiation pattern of the Rx-Ant, respectively. Moreover, $\theta_{n,r}^{\sss{\text{AOD}}}$ and $\varphi_{n,r}^{\sss{\text{AOD}}}$ are the elevation and azimuth AODs from $n^{th}$ RIS unit to the Rx-Ant, while $\theta_{n,r}^{\sss{\text{AOA}}}$ and $\varphi_{n,r}^{\sss{\text{AOA}}}$ denote the elevation and azimuth AOAs from $n^{th}$ RIS unit to the Rx-Ant.

    \section{Downlink Ergodic Capacity}
        Based on the above model, the equivalent channel between the Tx-Ant and Rx-Ant via $n^{th}$ RIS unit can be described as
            \begin{equation} \label{eq:channel-gn}
            \begin{aligned}
                \widehat{g}_{n}&=g_{n}z_{n}\mathcal{T}_{n},
            \end{aligned}
            \end{equation}
        where $\mathcal{T}_{n}$ denotes the reflection coefficient introduced by $n^{th}$ RIS unit. Usually, it is supposed that all the RIS units hold the same reflection amplitude $\mathcal{T}$ but individual phase shift $\phi_{n}$. Hence, the reflection coefficient can be rewritten as $\mathcal{T}_{n}=\mathcal{T}e^{-j\phi_{n}}$ with $\left|\mathcal{T}\right|\leq 1$ and $0\leq\phi_{n}\leq 2\pi$ \cite{YGaoDistrRISMISO,SZhangRISCapacityDeployment,SZengRISOrientaLocation}.

        On condition of the optimal phase adjustment at the RIS (the phase shift design of the RIS is beyond the scope of this paper because it has been investigated in a plethora of literature \cite{YChengPhaseAdjustmentIRS,SAbeywickramaRISPracticaPhase,CHuangRISDeepReinforcement,BDiRISHybridBeamforming}), the SNR at the Rx-Ant can be formulated as
            \begin{equation} \label{eq:SNR-max}
            \begin{aligned}
                \mathcal{SNR}&=\frac{\mathcal{T}^{2}P_{\text{t}}}{\mathcal{N}_{0}}\left(\sum_{\substack{n=1}}^{N}\left|g_{n}\right|\left|z_{n}\right|\right)^{2},
                          %&=\sum_{n=1}^{N}\left|h_{n}\right|^{2}\left|z_{n}\right|^{2}+\sum_{n=1}^{N}\sum_{\substack{i=1\\i\neq n}}^{N}\left|h_{n}\right|\left|z_{n}\right|\left|h_{i}\right|\left|z_{i}\right|
            \end{aligned}
            \end{equation}
        where $P_{\text{t}}$ and $\mathcal{N}_{0}$ denote the transmit power at the Tx-Ant and the noise power at the Rx-Ant, respectively.
        Furthermore, the downlink ergodic capacity is given by
            \begin{equation} \label{eq:capacity-downlink}
            \begin{aligned}
                C&=\mathbb{E}\left\{\log_{2}\left(1+\mathcal{SNR}\right)\right\}.
            \end{aligned}
            \end{equation}

        Unfortunately, the accurate closed-form expression seems out of reach, since the distribution of $\mathcal{SNR}$ in (\ref{eq:SNR-max}) is intractable. Some researchers have proposed approximate approaches to tackle this similar issue. For instance, leveraging the random matrix theory, \cite{COuyangergodiccapacityRISMIMO} presented the expressions of the upper and lower bounds of ergodic capacity by the channel dependent eigenvalues. In addition, \cite{QTaoRISPerformanceSISORician} and \cite{AMSalhabPerformanceRISRician} provided an asymptotic analysis of the upper bound of ergodic or average capacity using the Laguerre polynomial. In this paper, we provide a tight upper bound of the ergodic capacity in closed-form by generalized Hypergeometric function and it can be applicable for the RIS of any scale. It is summarized in the following.
        \begin{theorem}\label{theorem:capacity-upperbound}
            In the far-field of the RIS, assuming perfect reflection of all RIS units, i.e. $\left|\mathcal{T}\right|=1$, the upper bound of the ergodic capacity in downlink is given by
        \begin{equation}\label{eq:capacity-downlink-upbound}
        \begin{aligned}
            \overline{C}=\log_{2}\left(1+\frac{P_{\text{t}}}{\mathcal{N}_{0}}\rho_{\text{cc}}N\left[1+\left(N-1\right)\gamma\right]\right),
        \end{aligned}
        \end{equation}
        \end{theorem}
        where
        \begin{equation}\label{eq:gamma}
        \begin{aligned}
            \gamma=\frac{\pi^{2}{_{1}F_{1}}^{2}\left(\frac{3}{2};1;K_{1}\right){_{1}F_{1}}^{2}\left(\frac{3}{2};1;K_{2}\right)}{16\left(1+K_{1}\right)\left(1+K_{2}\right)e^{2\left(K_{1}+K_{2}\right)}},
        \end{aligned}
        \end{equation}
        and $\rho_{\text{cc}}$ denotes the composite channel gain (CCG) calculated by $\rho_{\text{cc}}=\rho_{t,c}\rho_{c,r}$ with
            \begin{small}
            \begin{equation} \label{eq:channel-gain-BS-RISc}
            \begin{aligned}
                 \rho_{t,c}&=\frac{\rho_{0}D_{t}D_{u}}{d_{t,c}^{2}}\left|F_{t}\left(\theta_{t,c}^{\sss{\text{AOD}}},\varphi_{t,c}^{\sss{\text{AOD}}}\right)\right|^{2}\left|F_{u}\left(\theta_{t,c}^{\sss{\text{AOA}}},\varphi_{t,c}^{\sss{\text{AOA}}}\right)\right|^{2},
            \end{aligned}
            \end{equation}
            \end{small}
            \begin{small}
            \begin{equation} \label{eq:channel-gain-RISc-Rx}
            \begin{aligned}
                 \rho_{c,r}&=\frac{\rho_{0}D_{u}D_{r}}{d_{c,r}^{2}}
                 \left|F_{u}\left(\theta_{c,r}^{\sss{\text{AOD}}},\varphi_{c,r}^{\sss{\text{AOD}}}\right)\right|^{2}\left|F_{r}\left(\theta_{c,r}^{\sss{\text{AOA}}},\varphi_{c,r}^{\sss{\text{AOA}}}\right)\right|^{2}.
            \end{aligned}
            \end{equation}
            \end{small}Here, $d_{t,c}$ denotes the distance between the Tx-Ant and RIS center, whilst $d_{c,r}$ is the distance between the RIS center and Rx-Ant. Additionally, $\theta_{t,c}^{\sss{\text{AOD}}}$ and $\varphi_{t,c}^{\sss{\text{AOD}}}$ represent the elevation and azimuth AODs from the Tx-Ant to RIS center, while $\theta_{c,r}^{\sss{\text{AOD}}}$ and $\varphi_{c,r}^{\sss{\text{AOD}}}$ are the elevation and azimuth AODs from the RIS center to Rx-Ant. Moreover, $\theta_{t,c}^{\sss{\text{AOA}}}$ and $\varphi_{t,c}^{\sss{\text{AOA}}}$ represent the elevation and azimuth AOAs from the Tx-Ant to RIS center, whilst $\theta_{c,r}^{\sss{\text{AOA}}}$ and $\varphi_{c,r}^{\sss{\text{AOA}}}$ are the elevation and azimuth AOAs from the RIS center to Rx-Ant.
        \begin{IEEEproof}
            Please see appendix \ref{proof:capacity-upperbound-theory}.
        \end{IEEEproof}

\begin{figure}
        	\centering
        	\begin{minipage}{.45\linewidth}
        		\centering
        		\includegraphics[width=\linewidth]{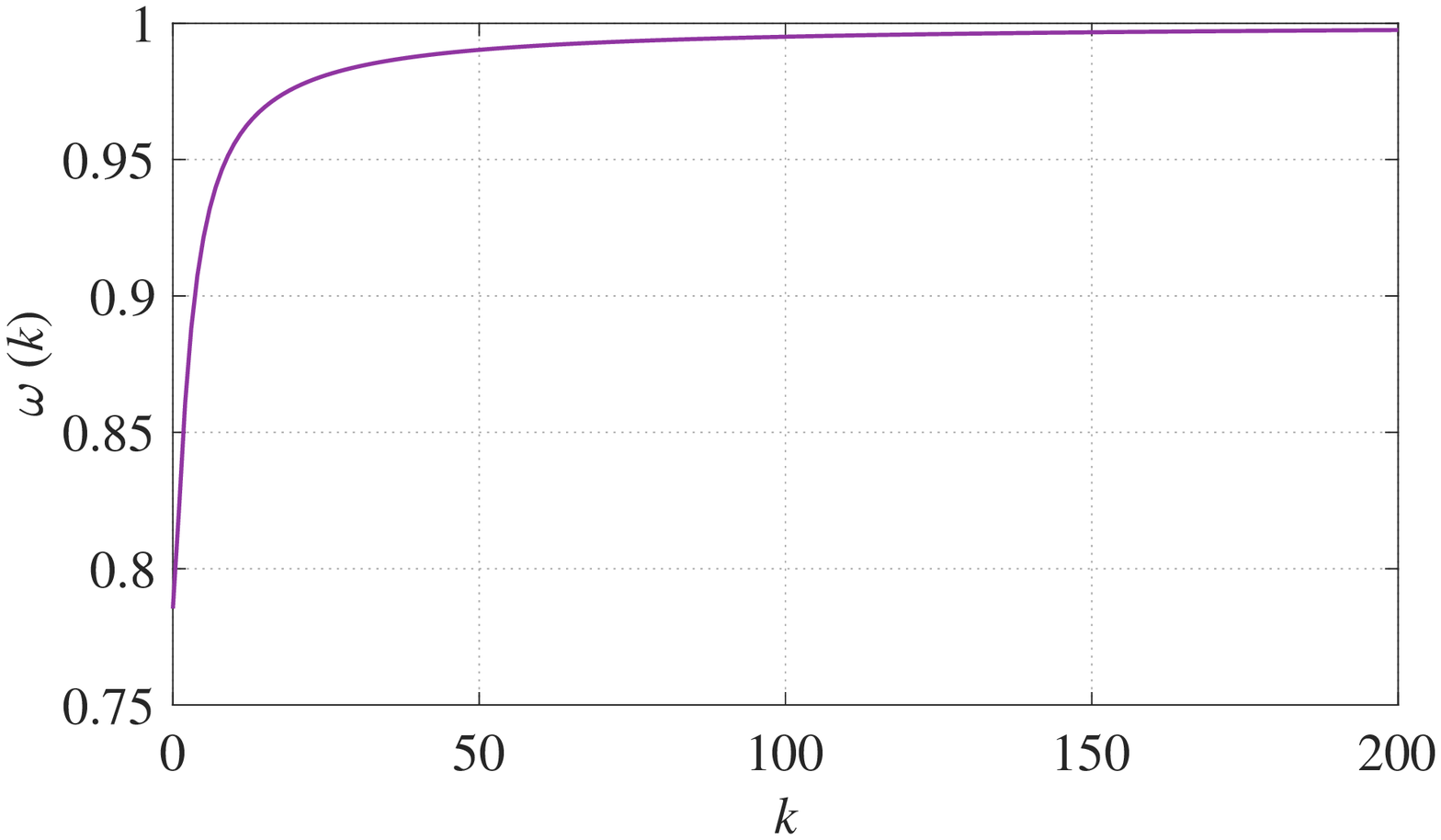}
        		\caption{$\omega \left(k\right)$ vs $k$.}
        		\label{Fig_2}
        	\end{minipage}
            \hspace{0.2in}
        	\begin{minipage}{.43\linewidth}
        		\centering
        		\includegraphics[width=\linewidth]{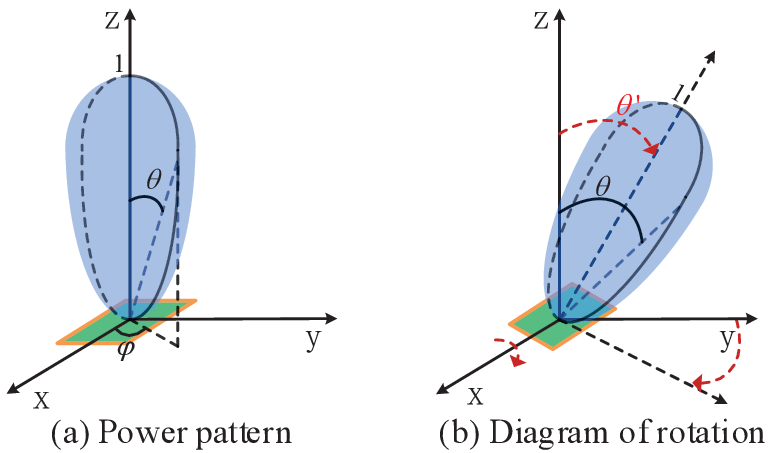}
        		\caption{Diagram of power pattern.}
        		\label{Fig_3}
        	\end{minipage}
        \end{figure}

        \begin{corollary}\label{corollary:capacity-K}
        To acquire deep insights on the ergodic capacity, we firstly investigate the properties of $\gamma$, involving the following cases.

        \begin{case1}\label{case:capacity-NLOS-NLOS}
        When $g_{n}$ and $z_{n}$ are Rayleigh channels, namely $K_1,K_2\rightarrow0$, $\gamma$ in (\ref{eq:gamma}) is given by
            \begin{equation}\label{eq:capacity-NLOS-NLOS}
            \begin{aligned}
                \gamma = \frac{\pi^{2}}{16}.
            \end{aligned}
            \end{equation}
        \end{case1}
        \begin{case1}\label{case:capacity-LOS-NLOS}
        When $g_{n}$ is a LOS link, i.e. $K_1\rightarrow\infty$, or $z_{n}$ is a LOS link, i.e. $K_2\rightarrow\infty$, $\gamma$ is given by
            \begin{equation}\label{eq:capacity-LOS-NLOS}
            \begin{aligned}
            \gamma = \left\{\begin{array}{ll}\frac{\pi{_{1}F_{1}}^{2}\left(\frac{3}{2};1;K_{2}\right)}{4\left(1+K_{2}\right)e^{2\left(K_{2}\right)}}, &K_1\rightarrow\infty,\\ \frac{\pi{_{1}F_{1}}^{2}\left(\frac{3}{2};1;K_{1}\right)}{4\left(1+K_{1}\right)e^{2\left(K_{1}\right)}}, & K_2\rightarrow\infty. \end{array}\right.
            \end{aligned}
            \end{equation}
        \end{case1}
        \begin{case1}\label{case:capacity-LOS-LOS}
        When $g_{n}$ and $z_{n}$ are both LOS links, i.e. $K_1\rightarrow\infty$ and $K_2\rightarrow\infty$, $\gamma$ can be simplified as
            \begin{equation}\label{eq:capacity-LOS-LOS}
            \begin{aligned}
                \gamma = 1.
            \end{aligned}
            \end{equation}
        \end{case1}
        \end{corollary}
        \begin{IEEEproof}
        Building on the series form of the Hypergeometric function \cite[Eq. (9.100)]{ISGTableofIntegrals}, we have
            \begin{equation}\label{eq:hyperfunc-limit}
                {_{1}F_{1}}\left(\frac{3}{2};1;k\right)=1+\frac{\frac{3}{2}}{1\cdot 1}k+\frac{\frac{3}{2}\left(\frac{3}{2}+1\right)}{1\cdot2\cdot 1\cdot 2}k^{2}+\frac{\frac{3}{2}\left(\frac{3}{2}+1\right)\left(\frac{3}{2}+2\right)}{1\cdot2\cdot 3\cdot 1\cdot 2\cdot 3}k^{3}+\dots
            \end{equation}
        Thus, when $k\rightarrow0$, we have $\lim_{k\rightarrow0}{_{1}F_{1}}\left(\frac{3}{2};1;k\right)=1$.
        For the sake of exposition, we define $\omega \left(k\right) = \pi{_{1}F_{1}}^{2}\left(\frac{3}{2};1;k\right)\left(1+k\right)^{-1}e^{-2k}/4$, then we have $\lim_{k\rightarrow0}\omega \left(k\right)=\pi/4$ definitely. Thus, $\gamma$ in (\ref{eq:gamma}) can be written as $\gamma= \omega \left(K_{1}\right) \omega \left(K_{2}\right)$, and we have $\gamma=\pi^{2}/16$ when $K_{1},K_{2}\rightarrow0$ as presented in \emph{Case \ref{case:capacity-NLOS-NLOS}}. Moreover, Fig. \ref{Fig_2} shows the value of $\omega \left(k\right)$ versus $k$. We have $\lim_{k\rightarrow\infty}\omega \left(k\right)=1$ ($k\in\left\{K_{1},K_{2}\right\}$) and then obtain the conclusions in \emph{Case \ref{case:capacity-LOS-NLOS}} and \emph{Case \ref{case:capacity-LOS-LOS}}.
        \end{IEEEproof}

        In light of \emph{Corollary \ref{corollary:capacity-K}}, it can be readily observed that $\gamma$ is a constant when $K_{1}$ and $K_{2}$ are certain. Especially, in the scenario dominated by LOS links, the SNR is proportional to $N^{2}$. Last but not the least, \emph{Theorem \ref{theorem:capacity-upperbound}} indicates that the capacity depends on the CCG, i.e. $\rho_{\text{cc}}$, which is not only sensitive to the rotations of the Tx-Ant, Rx-Ant and RIS, but also related to the distance between the Tx-Ant/Rx-Ant and RIS. Unfortunately, their relationships are not specific, so it deserves further analyses in the rest of this paper.

    \section{Optimization of Rotation Angle and Location}
        In this section, we devote our effort to investigate the optimal rotation angles of the Tx-Ant, Tx-Ant, and RIS at any location of the RIS, and then to explore the optimal location of the RIS.
        \subsection{Optimization of rotation angle}
            In this subsection, we focus on the investigations of the CCG by taking the radiation characteristics of antennas and RIS into account, and then extract the optimal rotation angles after analyzing the impact of rotation angles on the ergodic capacity. Let us kick off from the case without the rotation at the RIS, which is a common assumption in most of the literature.

            Before the discussion, let us revisit the basic definitions of the power pattern and directivity in antenna theory.
            \begin{definition}\label{def:power-pattern}
                Power Pattern: For an antenna with its major lobe directed along the Z-axis ($\theta=0$) as shown in Fig. \ref{Fig_3} (a), the normalized power pattern (or radiation intensity) is denoted as $U(\theta, \varphi)$, where $\theta$ and $\varphi$ represent the elevation and azimuth AODs from the antenna, respectively \cite{CABalanisAntennatheory,StutzmanAntennatheorydesign}. In practice, a general normalized power patterns can be typically represented by \cite[Eq. (2-31)]{CABalanisAntennatheory}
                    \begin{equation} \label{eq:power-pattern}
                        \begin{aligned}
                             U(\theta, \varphi)=\left\{\begin{array}{ll}\cos ^{q}(\theta), &\theta\in\left[0,\frac{\pi}{2}\right], \varphi\in\left[0,2\pi\right],\\ 0, & \text{elsewhere}, \end{array}\right.
                        \end{aligned}
                        \end{equation}
                    where $q$ is the parameter indicating the directivity. Please note that a larger $q$ may result in the stronger directivity but narrower pattern. Especially, $q=0$ implies an omnidirectional antenna.
            \end{definition}

            \begin{definition}\label{def:directivity}
                Directivity: The directivity of an antenna is defined as the ratio of the radiation intensity in a given direction from the antenna to the radiation intensity averaged over all directions \cite{CABalanisAntennatheory}. It can be tied more directly to the pattern function $D=4\pi/\iint\limits_{\Omega}U\left(\theta,\varphi\right)d_{\Omega}$, where $\Omega$ denotes the solid angle \cite{CABalanisAntennatheory,StutzmanAntennatheorydesign}.
            \end{definition}

        \subsubsection{Optimization of rotation angles at the antennas}
        When the antenna rotates around the X-axis with a degrees $\theta^{'}$ as depicted in Fig. \ref{Fig_3} (b), then the power pattern can be recast as
            \begin{small}
            \begin{equation} \label{eq:power-pattern-rotation}
            \begin{aligned}
                 U(\theta, \varphi)=\left\{\begin{array}{ll}\cos ^{q}(\theta-\theta^{'}), &\theta\in\left[\theta^{'},\theta^{'}+\frac{\pi}{2}\right], \varphi\in\left[0,2\pi\right],\\ 0, & \text{elsewhere}. \end{array}\right.
            \end{aligned}
            \end{equation}
            \end{small}Note that $\theta^{'}\in(0,\pi/2]$ denotes clockwise rotation and $\theta^{'}\in[-\pi/2,0)$ represents counterclockwise rotation.

        Hence, when the Tx-Ant and Rx-Ant rotate around the X-axis by $\theta_{t0}\in[-\pi/2,\pi/2]$ and $\theta_{r0}\in[-\pi/2,\pi/2]$, respectively, their power patterns can be recast as $U_{t}\left(\theta_{t}, \varphi_{t}\right)=\cos^{q_{t}}\left(\theta_{t}-\theta_{t\text{0}}\right)$ and $U_{r}\left(\theta_{r},\varphi_{r}\right)=\cos ^{q_{r}}\left(\theta_{r}-\theta_{r\text{0}}\right)$, correspondingly, where $q_{t}$ and $q_{r}$ are the directivity parameters of the Tx-Ant and Rx-Ant, respectively. As for the RIS without rotation, the power pattern of each unit can be illustrated as $U_{u}\left(\theta_{n}, \varphi_{n}\right)=\cos ^{q_{u}}\left(\theta_{n}\right)$ with $q_{u}$ being the directivity parameters of the RIS unit.

        As the power pattern in (\ref{eq:power-pattern}) adopted to model the Tx-Ant and Rx-Ant patterns, we can get the following theorem.
        \begin{theorem}\label{theorem:capacity-upperbound-tx-rx-rotaion}
            With only the rotations at the Tx-Ant and Rx-Ant, the CCG can be rewritten as
            \begin{equation} \label{eq:cascade-channel-gain-tx-rx}
            \begin{aligned}
                \rho_{\text{cc}}&=\frac{\rho_{0}^{2}10^{0.2\left(q_{t}+q_{r}+2q_{u}+4\right)}}{d_{t,c}^{2}d_{c,r}^{2}}\cos^{q_{t}}\left(\theta_{t,c}^{\sss{\text{AOD}}}-\theta_{t\text{0}}\right)\cos^{q_{u}}\left(\theta_{t,c}^{\sss{\text{AOA}}}\right)
                \cos^{q_{r}}\left(\theta_{c,r}^{\sss{\text{AOA}}}-\theta_{r\text{0}}\right)\cos^{q_{u}}\left(\theta_{c,r}^{\sss{\text{AOD}}}\right).
            \end{aligned}
            \end{equation}
        \end{theorem}
        \begin{IEEEproof}
            Recalling \emph{Definition \ref{def:directivity}}, when $U_{t}\left(\theta_{t}, \varphi_{t}\right)=\cos ^{q_{t}}\left(\theta_{t}\right)$, we can get the maximum directivity of the Tx-Ant $D_{t}=4\pi/\int_{0}^{2\pi}\int_{0}^{\pi/2}\cos ^{q_{t}}\left(\theta_{t}\right)\sin\left(\theta_{t}\right)d_{\theta_{t}}d_{\varphi_{t}}=2\left(q_{t}+1\right)$. Analogously, we can also obtain $D_{r}=2\left(q_{r}+1\right)$ and $D_{u}=2\left(q_{u}+1\right)$. Please note that they are calculated in dB in this form. According to \cite[Eq. (2-119)]{StutzmanAntennatheorydesign}, the normalized power pattern is simply the square of its field pattern magnitude, i.e. $U\left(\theta,\phi\right)=\left|F\left(\theta, \phi\right)\right|^{2}$. Then, by substituting $D_{t}$, $D_{r}$ and $D_{u}$, (\ref{eq:channel-gain-BS-RISc}) and (\ref{eq:channel-gain-RISc-Rx}) can be respectively rewritten as
            $\rho_{t,c}=\rho_{0}10^{0.2\left(q_{t}+q_{u}+2\right)} d_{t,c}^{-2}\cos^{q_{t}}\left(\theta_{t,c}^{\sss{\text{AOD}}}-\theta_{t\text{0}}\right)\cos^{q_{u}}\left(\theta_{t,c}^{\sss{\text{AOA}}}\right)$ and $\rho_{c,r}=\rho_{0}10^{0.2\left(q_{r}+q_{u}+2\right)}d_{c,r}^{-2}\cos^{q_{r}}\left(\theta_{c,r}^{\sss{\text{AOA}}}-\theta_{r\text{0}}\right)\cos^{q_{u}}\left(\theta_{c,r}^{\sss{\text{AOD}}}\right)$, respectively. Hence, (\ref{eq:cascade-channel-gain-tx-rx}) can be obtained.
        \end{IEEEproof}
        \begin{corollary} \label{corollary:transceiver-rotation-opt}
            The $\rho_{\text{cc}}$ can reach the maximum while the components $\cos^{q_{t}}\left(\theta_{t,c}^{\sss{\text{AOD}}}-\theta_{t\text{0}}\right)$ and $\cos^{q_{r}}\left(\theta_{c,r}^{\sss{\text{AOA}}}-\theta_{r\text{0}}\right)$ are maximized. That is, the optimal directions of the Tx-Ant and Rx-Ant are both pointing to the RIS center, i.e. $\theta_{t\text{0}}^{\text{opt}}=\theta_{t,c}^{\sss{\text{AOD}}}$ and $\theta_{r\text{0}}^{\text{opt}}=\theta_{c,r}^{\sss{\text{AOA}}}$. Therefore, the CCG in this case can be maximized as
            \begin{equation} \label{eq:cascade-channel-tx-rx-max}
            \begin{aligned}
                 \rho_{\text{cc}}&=\frac{\rho_{0}^{2}10^{0.2\left(q_{t}+q_{r}+2q_{u}+4\right)}}{ d_{t,c}^{2}d_{c,r}^{2}}\cos^{q_{u}}\left(\theta_{t,c}^{\sss{\text{AOA}}}\right)\cos^{q_{u}}\left(\theta_{c,r}^{\sss{\text{AOD}}}\right).
            \end{aligned}
            \end{equation}
        \end{corollary}

        \begin{remark} \label{remark:RIS-rotation-performance}
            Even with the optimal rotation angles at the Tx-Ant and Rx-Ant, the performance degradation is still unavoidable. This is because it is difficult to ensure that both the Tx-Ant and Rx-Ant are in the good direction (main-lobe) of RIS. In other words, one of them may in the poor direction (side-lobe) of the RIS. Precisely, $\theta_{t,c}^{\sss{\text{AOA}}}\rightarrow \pi/2$ and $\theta_{c,r}^{\sss{\text{AOD}}}\rightarrow\pi/2$ will result in $\cos^{q}\left(\theta_{t,c}^{\sss{\text{AOA}}}\right)\rightarrow0$ and $\cos^{q}\left(\theta_{c,r}^{\sss{\text{AOD}}}\right)\rightarrow0$, respectively. Naturally, there will be $\rho_{\text{cc}}=0$ as long as one of the Tx-Ant and Rx-Ant is in the poor direction, i.e. $\cos^{q}\left(\theta_{t,c}^{\sss{\text{AOA}}}\right)=0$ or $\cos^{q}\left(\theta_{c,r}^{\sss{\text{AOD}}}\right)=0$. This reveals that the RIS orientation may have a significant impact on the performance. Hence, it is possible to obtain a considerable performance gain by rotating the RIS. Please note that in order to discuss the potential effect of RIS rotation, the optimal rotations at the Tx-Ant and Rx-Ant are considered hereafter.
        \end{remark}
        \subsubsection{Optimization of rotation angle at the RIS}
        \begin{figure}[h]
            \centering
            \subfigure[Diagram of RIS rotation.]{
                \centering
                \label{Fig_4_1}
                \includegraphics[width=0.45\linewidth]{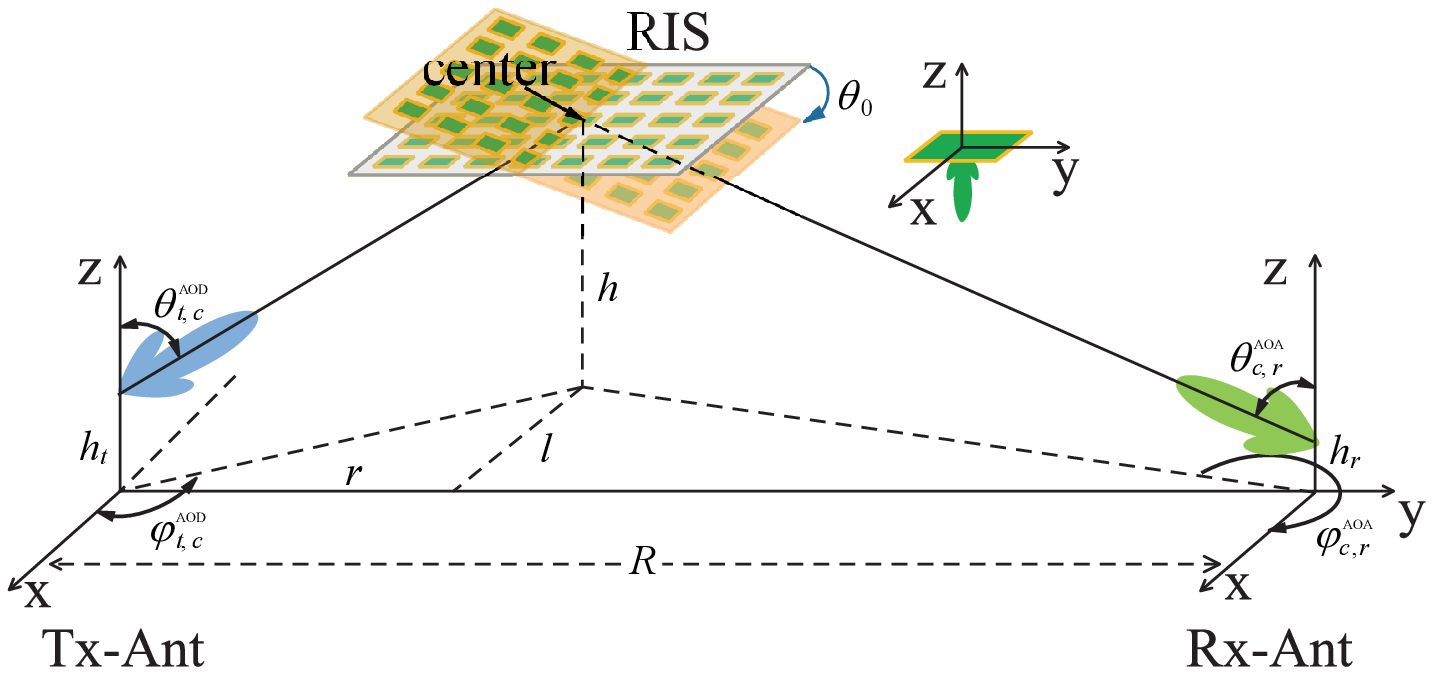}}
            \hspace{0.2in}
            \subfigure[Front view of the system.]{
                \centering
                \label{Fig_4_2}
                \includegraphics[width=0.45\linewidth]{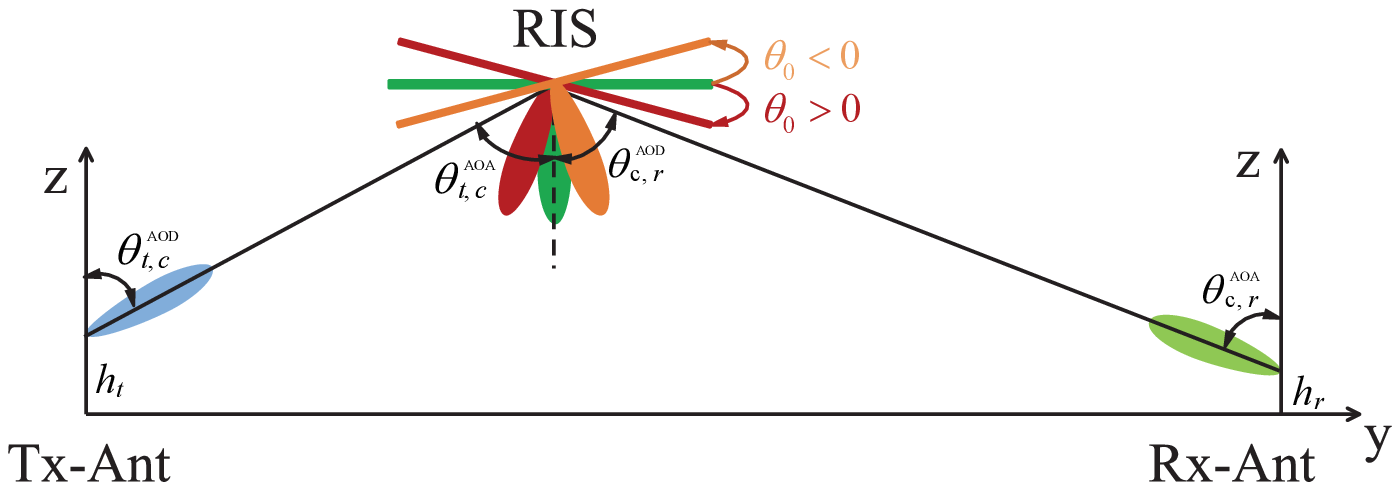}}
            \caption{Diagram of RIS-aided system with rotation.}
            \label{Fig_4}
        \end{figure}

        To describe the topology of the rotation at the RIS, Fig. \ref{Fig_4} \subref{Fig_4_1} pictorially depicts a diagram of RIS-aided system with rotation. As depicted, we denote $h_{t}$ and $h_{r}$ as the height of the Tx-Ant and Rx-Ant, respectively, and $h$ as the altitude of the RIS, without loss of generality. Additionally, we denote $R$ as the horizontal distance between the Tx-Ant and Rx-Ant, $l$ and $r$ as the vertical distance of the RIS to the Y-axis-Origin-Z-axis (YOZ) plane and that to the X-axis-Origin-Z-axis (XOZ) plane, respectively. Moreover, we define $\theta_{0}$ as the rotation angle around the X-axis, where $\theta_{0}>0$ and  $\theta_{0}<0$ represent clockwise and counterclockwise rotation, respectively. Defining $\alpha=\theta_{t,c}^{\sss{\text{AOA}}}$ and $\beta=\theta_{c,r}^{\sss{\text{AOD}}}$ for simplicity, we have the following properties.

        \begin{property} \label{property:alpha-beta-range}
%        \begin{equation} \label{eq:alpha-beta-range}
%            0<\alpha<\frac{\pi}{2},\quad 0<\beta<\frac{\pi}{2};
%        \end{equation}
        \begin{equation}\label{eq:alpha-beta-range}
            \begin{aligned}
                \left\{\begin{array}{l}0<\alpha<\frac{\pi}{2}\\ 0<\beta<\frac{\pi}{2}\\\alpha-\frac{\pi}{2}<\frac{\alpha-\beta}{2}<\frac{\pi}{2}-\beta
                \end{array}\right..
            \end{aligned}
        \end{equation}
%        \begin{equation} \label{eq:alpha-minus-beta}
%            \alpha-\frac{\pi}{2}<\frac{\alpha-\beta}{2}<\frac{\pi}{2}-\beta.
%        \end{equation}
        \end{property}
        \begin{IEEEproof}
            Based on the relative locations of the Tx-Ant/Rx-Ant and RIS in Fig. \ref{Fig_4} \subref{Fig_4_2}, it is clear that $0<\alpha<\pi/2$, $0<\beta< \pi/2$ and $\alpha+\beta<\pi$. Thus, we have $ 2\alpha-\pi<\alpha-\beta$, namely, $\alpha-\pi/2<(\alpha-\beta)/2$. Analogously, we have $\alpha-\beta<\pi-2\beta$, i.e.  $(\alpha-\beta)/2<\pi/2-\beta$. Therefore, the proof ends.
        \end{IEEEproof}

        \begin{property} \label{property:RIS-rotation-constrain}
            The rotation angle at the RIS constrains to
        \begin{equation} \label{eq:rotation-define-range}
            \alpha-\frac{\pi}{2}<\theta_{0}<\frac{\pi}{2}-\beta,
              \text{with}\left\{
            \begin{aligned}
                &\left[0,\frac{\pi}{2}-\beta\right), \;\text{clockwise},\\
                &\left(\alpha-\frac{\pi}{2},0\right], \;\text{counterclockwise}.
            \end{aligned}
            \right.
        \end{equation}
        \end{property}
        \begin{IEEEproof}
            This constraint is to ensure that the signal emitted from the Tx-Ant could reach the RIS and the reflecting signal by the RIS can arrive at the Rx-Ant. To be concrete, when the RIS rotates counterclockwise, i.e. $\theta_{0}<0$, the rotation angle cannot exceed $\left|\pi/2-\alpha\right|$, namely $\left|\theta_{0}\right|<\pi/2-\alpha$. Otherwise, the signal transmitted from the Tx-Ant cannot be reflected effectively by the RIS. Hence, we have $\theta_{0}>\alpha-\pi/2$. As for clockwise rotation, it is necessary to avoid a large rotation angle resulting the Rx-Ant being on the back of the RIS. Otherwise, the signal would be blocked, not enhanced. Therefore, we have $\left|\theta_{0}\right|<\pi/2-\beta$, i.e. $\theta_{0}<\pi/2-\beta$.
        \end{IEEEproof}

        Accordingly, besides the optimal rotation adjustment at the Tx-Ant/Rx-Ant, we provide the following theorem while considering the rotation at RIS.
        \begin{theorem} \label{theorem:cascade-channel-gain-rotation-max}
            To reap the best performance for the considered RIS-aided system, the optimal rotation angle at the RIS and the CCG are respectively given by
            \begin{equation} \label{eq:ris-rotation-opt}
            \begin{aligned}
                 \theta_{\text{0}}^{\text{opt}}=\frac{\theta_{t,c}^{\sss{\text{AOA}}}-\theta_{c,r}^{\sss{\text{AOD}}}}{2},
            \end{aligned}
            \end{equation}
            \begin{equation} \label{eq:cascade-channel-gain-rotation-max}
            \begin{aligned}
                 \rho_{\text{cc}}^{\text{R}}&=\frac{\rho_{0}^{2}10^{0.2\left(q_{t}+q_{r}+2q_{u}+4\right)}}{ d_{t,c}^{2}d_{c,r}^{2}}\frac{\left[1+\cos\left(\theta_{t,c}^{\sss{\text{AOA}}}+\theta_{c,r}^{\sss{\text{AOD}}}\right)\right]^{q_{u}}}{2^{q_{u}}}.
            \end{aligned}
            \end{equation}
        And, the ergodic capacity in this case can be described as
            \begin{equation}\label{eq:capacity-downlink-upbound-tx-rx-ris-rotaion}
                \begin{aligned}
                    \overline{C}_{\text{R}}=\log_{2}\left(1+\frac{P_{\text{t}}\rho_{0}^{2} 10^{0.2\left(q_{t}+q_{r}+2q_{u}+4\right)}\left[1+\cos\left(\theta_{t,c}^{\sss{\text{AOA}}}+\theta_{c,r}^{\sss{\text{AOD}}}\right)\right]^{q_{u}}}{\mathcal{N}_{0} d_{t,c}^{2}d_{c,r}^{2}2^{q_{u}}}N\left[1+\left(N-1\right)\gamma\right]\right).
                \end{aligned}
            \end{equation}
        \end{theorem}
       \begin{IEEEproof}
            Please see appendix \ref{proof:optimal-roration-theory}.
        \end{IEEEproof}
        \begin{remark} \label{remark:RIS-rotation-opt}
            \emph{Theorem \ref{theorem:cascade-channel-gain-rotation-max}} verifies that the appropriate rotation of RIS helps to avoid performance deterioration. Only when the RIS is located at the center of the Tx-Ant and Rx-Ant, it should be parallel to the XOY plane. When the RIS is close to the Tx-Ant, it should rotate  counterclockwise (towards the Rx-Ant) by $|\left(\theta_{t,c}^{\sss{\text{AOA}}}-\theta_{c,r}^{\sss{\text{AOD}}}\right)/2|$ and vice versa.
        \end{remark}
        \begin{remark} \label{remark:RIS-position-performance}
            Although we have obtained the optimal rotation at the RIS, the capacity is still related to the RIS location. Therefore, it is crucial to explore how to place the RIS reasonably with the optimal rotations.
        \end{remark}

    \subsection{Optimization of the RIS Location}
        As illustrated in Fig. \ref{Fig_4}, the Tx-Ant and Rx-Ant separate with a horizonal distance $R$ and vertical distance of the RIS to the XOZ plane $r\in\left[0,R\right]$. In addition, the RIS can move and rotate at an altitude $h\in\left[h_{\text{min}},h_{\text{max}}\right]$, where $h_{\text{min}}$ is the minimum altitude determined for safety reasons or to avoid obstacles, and $h_{\text{max}}$ is the maximum altitude regulated by law. Moreover, $l\in\left[0,L\right]$ denotes the vertical distance of the RIS to the YOZ plane. Our aim in this section is to evaluate the effect of the RIS location on the performance, including $r$, $h$, and  $l$, and then to explore the optimal location. We firstly present the location-dependent expressions as follows.
        \begin{theorem} \label{theorem:cascade-channel-gain-rotation-dist}
            With the optimal rotations at the Tx-Ant, Rx-Ant and RIS, the CCG dependents on the RIS location. And the location-dependent expression is given by
        \begin{small}
            \begin{equation}\label{eq:cascade-channel-gain-rotation-dist}
                \begin{aligned}
                    \rho_{\text{cc}}^{\text{R}}\left(l,r,h\right)&=\frac{\rho_{0}^{2}10^{0.2\left(q_{t}+q_{r}+2q_{u}+4\right)}}{2^{q_{u}}\left[r^{2}+l^{2}+\left(h-h_{t}\right)^{2}\right]\left[\left(R-r\right)^{2}+l^{2}+\left(h-h_{r}\right)^{2}\right]}\\
                 &\times\left[1+\frac{\left(h-h_{t}\right)\left(h-h_{r}\right)-\sqrt{\left(r^{2}+l^{2}\right)\left[\left(R-r\right)^{2}+l^{2}\right]}}{\sqrt{\left[r^{2}+l^{2}+\left(h-h_{t}\right)^{2}\right]\left[\left(R-r\right)^{2}+l^{2}+\left(h-h_{r}\right)^{2}\right]}}\right]^{q_{u}}.
                \end{aligned}
            \end{equation}
        \end{small}
        \end{theorem}
        \begin{IEEEproof}
            According to the geometric relationships, we have
            \begin{equation}\label{eq:distance-TxRISc-RIScRx}
                \begin{aligned}
                    \;\; \left\{\begin{array}{l}d_{t,c}=\sqrt{r^{2}+l^{2}+\left(h-h_{t}\right)^{2}}\\ d_{c,r}=\sqrt{\left(R-r\right)^{2}+l^{2}+\left(h-h_{r}\right)^{2}}
                    \end{array}\right.,
                \end{aligned}
            \end{equation}
            \begin{equation}\label{eq:varphi-TxRISc-AOA-RIScRx-AOD}
                \begin{aligned}
                    \left\{\begin{array}{l}\cos\left(\theta_{t,c}^{\sss{\text{AOA}}}\right)=\frac{h-h_{t}}{\sqrt{r^{2}+l^{2}+\left(h-h_{t}\right)^{2}}}\\
                    \cos\left(\theta_{c,r}^{\sss{\text{AOD}}}\right)=\frac{h-h_{r}}{\sqrt{\left(R-r\right)^{2}+l^{2}+\left(h-h_{r}\right)^{2}}}
                    \end{array}\right..
                \end{aligned}
            \end{equation}
            By substituting (\ref{eq:distance-TxRISc-RIScRx}) and (\ref{eq:varphi-TxRISc-AOA-RIScRx-AOD}) into (\ref{eq:cascade-channel-gain-rotation-max}), the ergodic capacity of the location function (\ref{eq:cascade-channel-gain-rotation-dist}) is obtained.
        \end{IEEEproof}

        \begin{corollary} \label{corollary:l-opt}
            The RIS ought to be preferentially deployed directly above the Tx-Rx pair, i.e. $l=0$, then (\ref{eq:cascade-channel-gain-rotation-dist}) can be recast as
            \begin{small}
            \begin{equation} \label{eq:cascade-channel-gain-l-opt}
            \begin{aligned}
                \rho_{\text{cc}}^{\text{R}}\left(r,h\right)&=\frac{\rho_{0}^{2}10^{0.2\left(q_{t}+q_{r}+2q_{u}+4\right)}}{2^{q_{u}}\left[r^{2}+\left(h-h_{t}\right)^{2}\right]\left[\left(R-r\right)^{2}+\left(h-h_{r}\right)^{2}\right]}\\
                 &\times\left[1+\frac{\left(h-h_{t}\right)\left(h-h_{r}\right)-r\left(R-r\right)}{\sqrt{\left[r^{2}+\left(h-h_{t}\right)^{2}\right]\left[\left(R-r\right)^{2}+\left(h-h_{r}\right)^{2}\right]}}\right]^{q_{u}}.
            \end{aligned}
            \end{equation}
            \end{small}
        \end{corollary}

          \begin{IEEEproof}
            According to the derivative of (\ref{eq:cascade-channel-gain-rotation-dist}) with respect to $l$, we find $\partial \rho_{\text{cc}}^{\text{R}}/\partial l<0$, which means that the CCG is monotonically decreasing as $l$ increases, omitted for simplify. Hence, $l=0$ is the optimal vertical distance. Then, substituting $l=0$ into (\ref{eq:cascade-channel-gain-rotation-dist}), (\ref{eq:cascade-channel-gain-l-opt}) is obtained.
        \end{IEEEproof}

        Usually, the Tx-Ant and Rx-Ant are placed on the ground and their heights are negligible compared to the altitude of the RIS. Therefore, when $l=0$, neglecting $h_{t}$ and $h_{r}$, (\ref{eq:distance-TxRISc-RIScRx}) and (\ref{eq:varphi-TxRISc-AOA-RIScRx-AOD}) can be simplified. Then, we have
            \begin{equation}
                \begin{aligned}
                \cos\left(\theta_{t,c}^{\sss{\text{AOA}}}+\theta_{c,r}^{\sss{\text{AOD}}}\right)
                &=\frac{r^{2}-Rr+h^{2}}{\sqrt{\left(r^{2}+h^{2}\right)\left[\left(R-r\right)^{2}+h^{2}\right]}}.
                \end{aligned}
            \end{equation}
        Hence, (\ref{eq:cascade-channel-gain-l-opt}) can be further approximated as
            \begin{equation} \label{eq:cascade-channel-gain-appro}
            \begin{aligned}
                \rho_{\text{cc}}^{\text{R}}\left(r,h\right)&=\frac{\rho_{0}^{2}10^{0.2\left(q_{t}+q_{r}+2q_{u}+4\right)}}{2^{q_{u}}\left(r^{2}+h^{2}\right)\left[\left(R-r\right)^{2}+h^{2}\right]}
                 \left[1+\frac{r^{2}-Rr+h^{2}}{\sqrt{\left(r^{2}+h^{2}\right)\left[\left(R-r\right)^{2}+h^{2}\right]}}\right]^{q_{u}}.
            \end{aligned}
            \end{equation}
        \begin{remark}\label{remark:cos-alpha-plus-beta}
            It is clear that $-1<\cos\left(\varphi_{t,c}^{\sss{\text{AOA}}}+\varphi_{c,r}^{\sss{\text{AOD}}}\right)<1$ due to the fact that $0<\varphi_{t,c}^{\sss{\text{AOA}}}+\varphi_{c,r}^{\sss{\text{AOD}}}<\pi$. Thus, for arbitrary $h\in\left[h_{\text{min}},h_{\text{max}}\right]$ and $r\in\left[0,R\right]$, we have
            \begin{equation} \label{eq:cos-alpha-plus-beta}
                1+\frac{r^{2}-Rr+h^{2}}{\sqrt{\left(r^{2}+h^{2}\right)\left[\left(R-r\right)^{2}+h^{2}\right]}}>0.
            \end{equation}
        \end{remark}

        Furthermore, revisiting the optimal rotation at the RIS in (\ref{eq:ris-rotation-opt}), its location-dependent expression is given by
          \begin{equation} \label{eq:ris-rotation-opt-location-dep}
            \begin{aligned}
                 \theta_{\text{0}}^{\text{opt}}=\frac{\arccos\frac{h}{\sqrt{r^{2}+h^{2}}}-\arccos\frac{h}{\sqrt{\left(R-r\right)^{2}+h^{2}}}}{2}.
            \end{aligned}
            \end{equation}

        \begin{corollary} \label{corollary:r-opt}
        For deep insights into the impact of the RIS location on the ergodic capacity, we have the following cases.
        \begin{case2} \label{case:R-largethan-sqrtl2h2}
              When the RIS is deployed with a low altitude, i.e. $h<R/2$, the optimal value of $r$ is in the interval $r^{\text{opt}}\in\left[0,(R-\sqrt{R^{2}-4h^{2}})/2\right]$ or $r^{\text{opt}}\in\left[(R+\sqrt{R^{2}-4h^{2}})/2,R\right]$.
        \end{case2}
        \begin{case2} \label{case:R-muchlargethan-sqrtl2h2}
         When $h\ll R/2$, the RIS is preferably deployed directly above the Tx-Ant or Rx-Ant, i.e. $r^{\text{opt}}=0$ or $r^{\text{opt}}=R$, and the maximum CCG in this case is given by
            \begin{equation} \label{eq:cascade-channel-gain-rotation-dist-max_specialcase}
            \begin{aligned}
                \rho_{\text{cc}}^{\text{R}}\left(r^{\text{opt}},h\right)&=\frac{\rho_{0}^{2}10^{0.2\left(q_{t}+q_{r}+2q_{u}+4\right)}}{2^{q_{u}} h^{2}\left(R^{2}+h^{2}\right)}
                \left[1+\frac{h}{\sqrt{\left(R^{2}+h^{2}\right)}}\right]^{q_{u}}.
            \end{aligned}
            \end{equation}
        \end{case2}
        \begin{case2} \label{case:R-lessthan-sqrtl2h2}
            When $h>R/2$, it is difficult to obtain the optimal closed-from expression of $r$, as well the optimal interval.
        \end{case2}
        \end{corollary}
        \begin{IEEEproof}
            See Appendix \ref{proof:r-optimal}.
        \end{IEEEproof}

        In order to get the optimal $h$, we firstly take the derivatives of $\mu\left(r,h\right)$ and $\nu\left(r,h\right)$ with respect to $h$. Please note that the $\mu\left(r,h\right)$ and $\nu\left(r,h\right)$ are defined in Appendix \ref{proof:r-optimal}. The derivations are respectively given by
            \begin{small}
            \begin{equation} \label{eq:gl-derivation}
            \begin{aligned}
                \frac{\partial \mu}{\partial h}&=\frac{q_{u}R^{2}h\left(h^{2}+Rr-r^{2}\right)}{\left(r^{2}+h^{2}\right)^{\frac{3}{2}}\left[\left(R-r\right)^{2}+h^{2}\right]^{\frac{3}{2}}}
                \left[1+\frac{r^{2}-Rr+h^{2}}{\sqrt{\left(r^{2}+h^{2}\right)\left[\left(R-r\right)^{2}+h^{2}\right]}}\right]^{q_{u}-1},
            \end{aligned}
            \end{equation}
            \end{small}
            \begin{small}
            \begin{equation} \label{eq:hl-derivation}
            \begin{aligned}
                \frac{\partial \nu}{\partial h}=-\frac{2h\left[2h^{2}+r^{2}+\left(R-r\right)^{2}\right]}{\left(r^{2}+h^{2}\right)^{2}\left[\left(R-r\right)^{2}+h^{2}\right]^{2}}.
            \end{aligned}
            \end{equation}
            \end{small}

        Building on the insights from \emph{Remark \ref{remark:cos-alpha-plus-beta}}, it is obvious that $\partial \mu/\partial h>0$ and $\partial \nu/\partial h<0$ for arbitrary $h$. This indicates that $\mu\left(r,h\right)$ increases monotonically, while $\nu\left(r,h\right)$ decreases monotonically for arbitrary $h\in\left[h_{\text{min}},h_{\text{max}}\right]$. Hence, the effect of $h$ on the CCG is uncertain, which implies the impossibility of extracting the optimal $h$ theoretically.
             \begin{algorithm}
            	\label{alg:optimal-position}
            	\caption{\small Algorithm for the optimal rotation and location.}
            	\small
            	\LinesNumbered
            	\KwIn{\small Initial key parameters: $N$, $P_{t}$, $q_{t}$, $q_{r}$, $q_{u}$, $\lambda$, $K_1$, $K_2$, $\mathcal{N}_{0}$, $R$, $h_{\text{min}}$, $h_{\text{max}}$}
            	\KwOut{\small The optimal RIS location $(r^{\text{opt}}, h^{\text{opt}})$, optimal RIS rotation ($\theta_{\text{0}}^{\text{opt}}$)}
                \ForEach{$h\in \left[h_{\text{min}},h_{\text{max}}\right]$}
                {
                    {\bf{Initialize}} $r = 0$\;
                    Calculate $\Delta=R^{2}-4 h^{2}$\;
                    \eIf{$\Delta>0$}
                    {
                		Set $r_{\text{max}}=\frac{R-\sqrt{\Delta}}{2}$\;
                    }
                    {
                        Set $r_{\text{max}}=\frac{R}{2}$\;
                    }
                    \ForEach{$r\in\left[0,r_{\text{max}}\right]$}
                    {
                	   Calculate $\rho^{\text{R}}_{\text{cc}}\left(r,h\right)$ according to (\ref{eq:cascade-channel-gain-appro}), and find the maximum gain\;
                	}
                    Calculate and compare gains to find the maximum one $\rho^{\text{R}}_{\text{cc}}\left(r^{\text{opt}},h\right)$\;
                }
                Compare gains $\rho^{\text{R}}_{\text{cc}}\left(r^{\text{opt}},h\right)$ and find the maximum one $\rho^{\text{R}}_{\text{cc}}\left(r^{\text{opt}},h^{\text{opt}}\right)$\;
                Obtain the optimal location $\left(r^{\text{opt}}, h^{\text{opt}}\right)$ and $\left(R-r^{\text{opt}}, h^{\text{opt}}\right)$\;
                Calculate the optimal RIS rotation $\theta_{\text{0}}^{\text{opt}}$ according to (\ref{eq:ris-rotation-opt-location-dep}).
            \end{algorithm}
        \subsubsection{Optimal location selection}
            In view of the above discussions, it is tough to obtain the optimal RIS location theoretically, since the effects of $r$ and $h$ are intractable. In order to find the optimal location, we propose a simplified exhaustive algorithm as presented in \emph{Algorithm 1}.

            Due to the uncertain effect of $h$, it is necessary to go through all its values. To avoid the calculation of the redundant values of $r$ and reduce the computation time, we narrow the valid range of $r$ by comparing $\Delta$ ($\Delta=R^{2}-4 h^{2}$) with $0$. Then, we can get the optimal $r$ for each $h$, as well the maximum channel gain. After calculating the ergodic values of $h$, we can obtain the maximum gain and optimal location, i.e. $\left(r^{\text{opt}},h^{\text{opt}}\right)$. Note that, due to the symmetry of the CCG with respect to $r=R/2$, the location $\left(R-r^{\text{opt}},h^{\text{opt}}\right)$ can be also adopted as the optimal location. Accordingly, the optimal rotation of the RIS can be calculated according to (\ref{eq:ris-rotation-opt-location-dep}).
            \begin{algorithm}
            	\label{alg:effective-area}
            	\caption{\small Algorithm for the effective region of RIS deployment.}
            	\small
            	\LinesNumbered
            	\KwIn{\small Initial key parameters: $N$, $P_{t}$, $q_{t}$, $q_{r}$, $q_{u}$, $\lambda$, $K_1$, $K_2$, $\mathcal{N}_{0}$, $R$, $h_{\text{min}}$, $h_{\text{max}}$}
            	\KwOut{\small The effective region}
                \ForEach{$h \in \left[h_{\text{min}},h_{\text{max}}\right]$}
                {
                    {\bf{Initialize}} $r = 0$\;
                    Calculate $\Delta=R^{2}-4h^{2}$\;
                    \eIf{$\Delta>0$}
                    {
                		Set $r=\frac{R-\sqrt{\Delta}}{2}$\;
                        \eIf{$\rho^{\text{R}}_{\text{cc}}\left(r,h\right)\geq\frac{\mathcal{N}_{0}\gamma_{\text{th}}}{NP_{t}\left[1+\left(N-1\right)\gamma\right]}$}
                        {
                            Set $r_{1}=\frac{R-\sqrt{\Delta}}{2}$\;
                            \ForEach{$r\in\left[0,r_{1}\right]$}
                                {\small{Find the set $\mathcal{A}_{1}$ satisfies (\ref{eq:effective-area-selection})}\;}
                            Set $t=1$\, $r_{2}=r_{1}+\frac{\sqrt{\Delta}}{2^{t+1}}$\;
                                \eIf{$\rho^{\text{R}}_{\text{cc}}\left(r_{2},h\right)<\frac{\mathcal{N}_{0}\gamma_{\text{th}}}{NP_{t}\left[1+\left(N-1\right)\gamma\right]}$}
                                {
                                    \Repeat{$\rho^{\text{R}}_{\text{cc}}\left(r_{2},h\right)\geq\frac{\mathcal{N}_{0}\gamma_{\text{th}}}{NP_{t}\left[1+\left(N-1\right)\gamma\right]}$}
                                    {
                                        Update $t=t+1$, $r_{2}=r_{1}+\frac{\sqrt{\Delta}}{2^{t+1}}$\;
                                    }
                                }
                                {
                                    \Repeat{$\rho^{\text{R}}_{\text{cc}}\left(r_{2},h\right)\leq\frac{\mathcal{N}_{0}\gamma_{\text{th}}}{NP_{t}\left[1+\left(N-1\right)\gamma\right]}$}
                                    {
                                        Update $t=t+1$, $r_{2}=\frac{R}{2}-\frac{\sqrt{\Delta}}{2^{t+1}}$\;
                                    }
                                }
                                Obtain the optional region $\left[r_{1},r_{2}\right]$\;
                        }
                        {
                            Set $r_{1}=\frac{R-\sqrt{\Delta}}{2}$\;
                            \ForEach{$r\in\left[0,r_{1}\right]$}
                                {Find the set $\mathcal{A}_{2}$ satisfies (\ref{eq:effective-area-selection})\;}
                        }
                        Obtain the optional $r$: $\mathcal{A}_{1}\cup\left[r_{1},r_{2}\right]$ or $\mathcal{A}_{2}$\;
                    }
                    {
                        \ForEach{$r\in\left[0,\frac{R}{2}\right]$}
                            {Find the set $\mathcal{A}_{3}$ satisfies (\ref{eq:effective-area-selection})\;}
                    }
                    Obtain optional $r$: $\mathcal{A}_{1}\cup\left[r_{1},r_{2}\right]$ or $\mathcal{A}_{2}$ or $\mathcal{A}_{3}$ for each $h$.
                }
            \end{algorithm}

        \subsubsection{Effective regions selection}
        Instead of choosing the optimal location of the RIS, it would be more attractive to figure out a region, where the RIS can be deployed flexibly to enhance the communications in practice. For example, the RIS cannot be deployed at the theoretical optimal locations in the realistic scene because those locations will result in blocking or have been occupied by some existing infrastructures.

        For these purposes, we migrate towards finding out the effective regions, which are defined as follows.
        \begin{definition}
         Effective regions denote the positions or areas that are suitable for RIS deployment under the condition that the SNR is not less than a given threshold $\gamma_{\text{th}}$, i.e. $\mathbb{E}\left\{\mathcal{SNR}\right\}\geq\gamma_{\text{th}}$.
        \end{definition}

        According to (\ref{eq:cascade-channel-gain-appro}) and (\ref{eq:SNR-max-Exp}), we have $\mathbb{E}\left\{\mathcal{SNR}\right\}=P_{\text{t}}\mathcal{N}_{0}^{-1}N\rho^{\text{R}}_{\text{cc}}\left(r,h\right)\left[1+\left(N-1\right)\gamma\right]$. Then, the effective regions should satisfy that
            \begin{equation} \label{eq:effective-area-selection}
            \begin{aligned}
                     \rho^{\text{R}}_{\text{cc}}\left(r,h\right)\geq\frac{\mathcal{N}_{0}\gamma_{\text{th}}}{NP_{t}\left[1+\left(N-1\right)\gamma\right]}.
            \end{aligned}
            \end{equation}

        To figure out the effective regions, we provide an algorithm as described in \emph{Algorithm 2}. It is necessary to go through all the values of $h$ and $r$ when $r\in\left[0,r_{1}\right]$ due to the uncertain effect of them. As for the interval $r\in\left[r_{1},R/2\right]$, when $\Delta>0$, a bisection method is adopted to determine the critical value of $r$ quickly because it is monotonically decreasing in this interval.

        In addition, revisiting (\ref{eq:cascade-channel-tx-rx-max}), the location-dependent expression of the CCG without rotation at the RIS is given by
            \begin{equation} \label{eq:cascade-channel-gain-dist-max}
            \begin{aligned}
                     \rho_{\text{cc}}\left(r,h\right)&=\frac{\rho_{0}^{2}10^{0.2\left(q_{t}+q_{r}+2q_{u}+4\right)}}{\left(r^{2}+l^{2}+h^{2}\right)\left[\left(R-r\right)^{2}+l^{2}+h^{2}\right]}\\
                     &\times\left[\frac{h^{2}}{\sqrt{\left(r^{2}+l^{2}+h^{2}\right)\left[\left(R-r\right)^{2}+l^{2}+h^{2}\right]}}\right]^{q_{u}}.
            \end{aligned}
            \end{equation}
        This would be presented as a benchmark, in this paper, helping to grasp the effectiveness of the RIS rotation adjustment on the performance improvement.
    \section{Simulation Results}
        In this section, simulation results are presented to validate the theoretical analysis of the ergodic capacity, the impact of the antennas and RIS rotations, and the effectiveness of the proposed RIS deployment. The simulations are based on the scenario illustrated in Fig. \ref{Fig_1} and Fig. \ref{Fig_4}. The key parameters are listed in Table \ref{tab:parameters} as follows.
        \begin{table}[!h]
        	\centering
        	\caption{List of parameters.}
            \renewcommand\arraystretch{0.8}
          	\label{tab:parameters}
        	%\begin{tabularx}{0.8\textwidth}{c|X}
            \begin{tabular}{m{1.5cm}<{\centering}|m{2cm}<{\centering}||m{1.5cm}<{\centering}|m{2cm}<{\centering}||m{1.5cm}<{\centering}|m{2cm}<{\centering}}
        		\hline
                \hline
        		\bf{Parameters}    & \bf{Values}      & \bf{Parameters}      & \bf{Values}   & \bf{Parameters}   & \bf{Values} \\
                \hline
        		$N$                & $16\sim 128$     & $B$                  & 5MHz          & $l$               & 100m \\
        		\hline
        		$K_{1}$            & 5                & $K_{2}$              &5              & $\rho_{0}$        &-40 dBm\\
                \hline
        		$\lambda$          & 0.125m           & $R$                  & 1000m         & $q_{t}$           & 20 \\
        		\hline
        		$\sigma$           & $-174$dBm/Hz     & $h_{\text{min}}$     & 100m          & $q_{r}$           & 20 \\
                \hline
        		$P_{\text{t}}$     & $-10\sim 30$dBm  & $h_{\text{max}}$     & 600m          & $q_{u}$           & $2\sim6$ \\
        		\hline
                \hline
        	\end{tabular}
\end{table}

        For simplicity, the RIS location is denoted by $P(l,r,h)$. For example, the initial RIS location is assumed as $l=100$, $r=200$, and $h=100$, hence it is denoted by $P(100,200,100)$. Additionally, $q_{u}=4$, $N=64$, and $P_{t}=10$ dBm unless otherwise stated. Moreover, the initial directions of the Tx-Ant and Rx-Ant are in the positive direction of the Z-axis, whilst the initial direction of the RIS points directly to the negative direction of the Z-axis. In other words, there is no rotation at the Tx-Ant, Rx-Ant, and RIS initially, i.e. $\theta_{t0}=0^{\circ}$, $\theta_{r0}=0^{\circ}$, and $\theta_{0}=0^{\circ}$, respectively. Furthermore, the noise power $\mathcal{N}_{0} =\sigma B$ with $\sigma$ and $B$ being the noise density and bandwidth, respectively.
            \begin{figure}
        	\centering
        	\begin{minipage}{.45\linewidth}
                \centering
                \includegraphics[width=\linewidth]{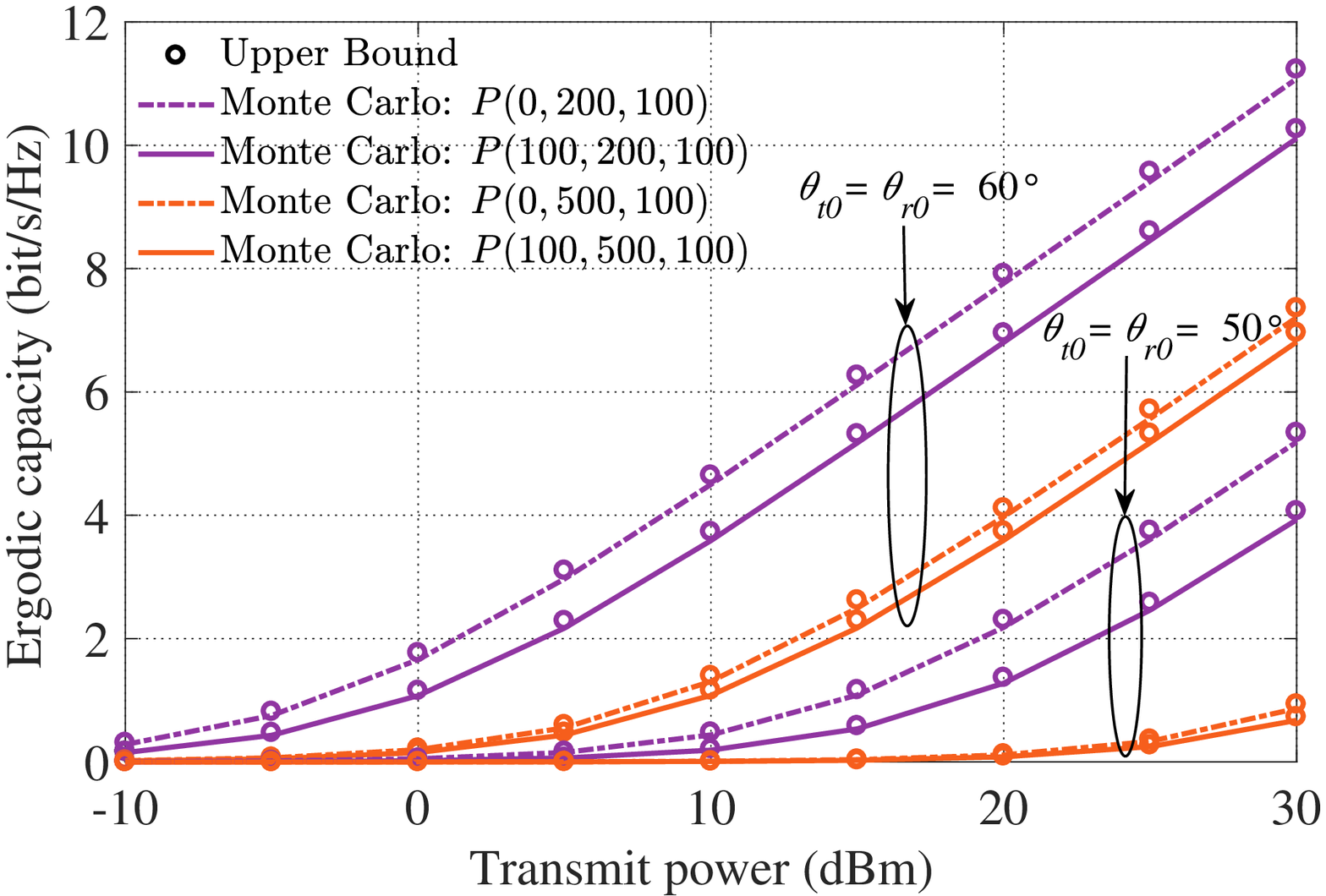}
                \caption{The upper bounds and the Monte Carlo results of the ergodic capacity.}
                \label{Fig_5}
        	\end{minipage}
           \hspace{0.2in}
        	\begin{minipage}{.45\linewidth}
                \centering
                \includegraphics[width=\linewidth]{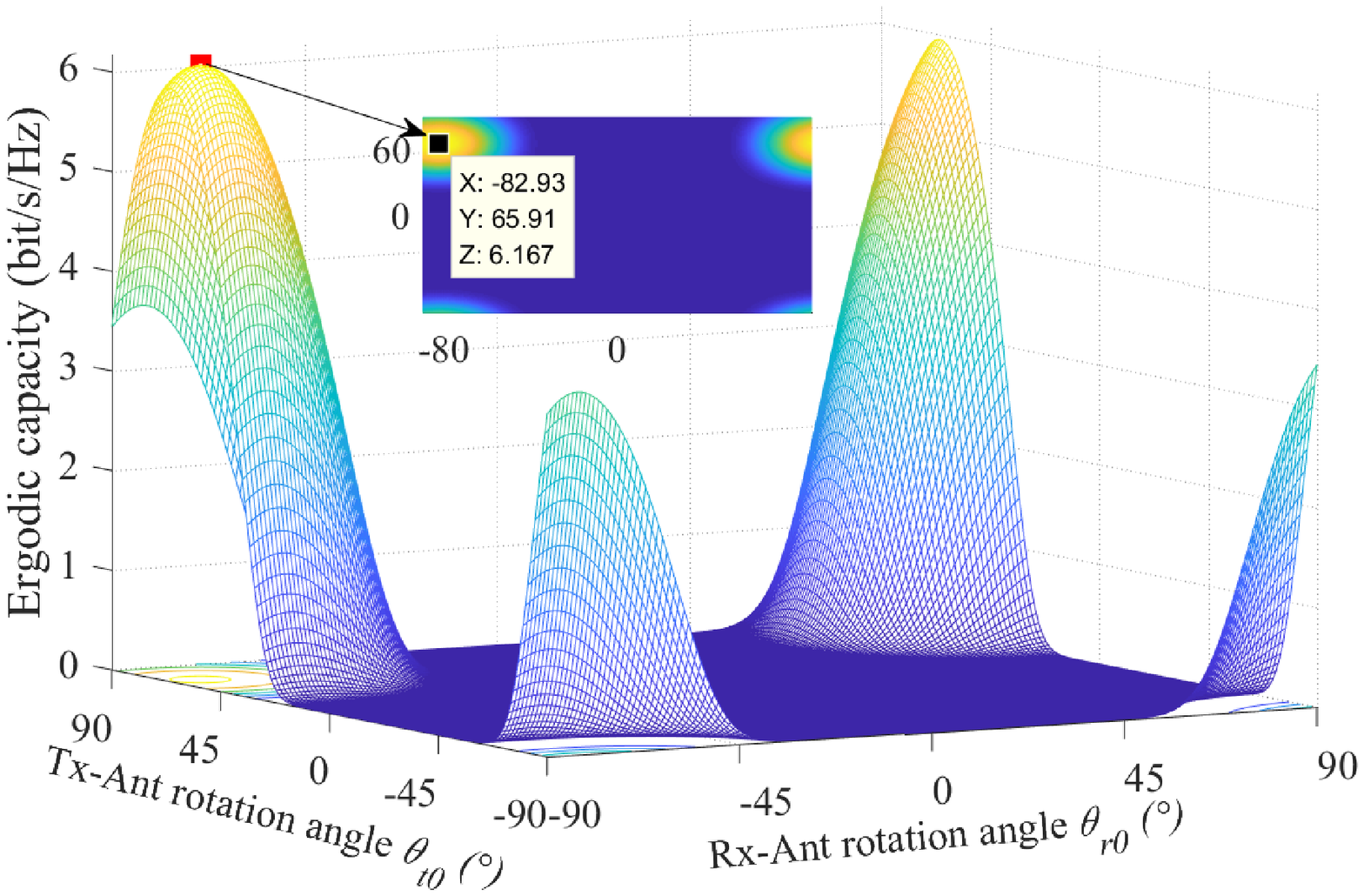}
                \caption{Impact of $\theta_{t0}$ and $\theta_{r0}$ on the ergodic capacity.}
                \label{Fig_6}
        	\end{minipage}
            \end{figure}

        \subsection{Impact of the rotation angles on the capacity}
            We firstly investigate the tightness of the theoretical upper bound of the ergodic capacity. As depicted in Fig. \ref{Fig_5}, we exemplify two cases, including $\theta_{t0}=\theta_{r0}=50^{\circ}$ and $\theta_{t0}=\theta_{r0}=60^{\circ}$, of four different locations, i.e. $P(0,200,100)$, $P(100,200,100)$, $P(0,500,100)$, and $P(100,500,100)$. Note that, there is no rotation at the RIS in these cases. It is clear that the upper bound is tightly closed with the Monte Carlo results. This suggests the accuracy of the theoretical expressions in \emph{Theorem \ref{theorem:capacity-upperbound}} and \emph{Theorem \ref{theorem:capacity-upperbound-tx-rx-rotaion}}.

            Besides, we notice that the antenna rotation angles have a huge impact on the capacity. More specifically, improper antenna orientation will incur a serious performance deterioration. For example, as illustrated in the purple solid line, i.e. $P(100,200,100)$, there is a big gap between the capacity with $50^{\circ}$ rotation and that with $60^{\circ}$ rotation. Also, this phenomenon exists in the other configurations. Furthermore, $r$ and $l$ have a significant impact on the performance. This indicates the importance of the RIS location selection.
%        \begin{figure}[!t]
%            \centering
%            \includegraphics[width=0.85\linewidth]{./Figs/TxRx_Rotation.eps}
%            \caption{ Impact of Tx-Ant/Rx-Ant rotation on the ergodic capacity.}
%            \label{fig:Capacity-rotation-txrx}
%        \end{figure}

            Fig. \ref{Fig_6} illustrates the impact of the rotation angles at the Tx-Ant ($\theta_{t0}$) and that at the Rx-Ant ($\theta_{r0}$) on the ergodic capacity. In order to elaborate the influence specifically, we consider a certain location, $P(100,200,100)$ as an example. Based on the geometric relationships, we have $\theta_{t,c}^{\sss{\text{AOA}}}=65.9052^{\circ}$ and $\theta_{c,r}^{\sss{\text{AOA}}}=-82.9294^{\circ}$ in this case. As can be observed, the maximum capacity achieved with $65.91^{\circ}$ rotation at the Tx-Ant and $-82.93^{\circ}$ rotation at the Rx-Ant, indicating the correctness of \emph{Corollary \ref{corollary:transceiver-rotation-opt}}. Please note that the minus sign represents counterclockwise rotation. Hereafter, the optimal rotations at the Tx-Ant and Rx-Ant are adopted in the following discussions unless otherwise stated.

    %        \begin{figure}[!t]
    %            \centering
    %            \includegraphics[width=0.85\linewidth]{./Figs/ErgodiCapacity_qu.eps}
    %            \caption{Impact of $q_u$ on the ergodic capacity.}
    %            \label{fig:Capacity-qu}
    %        \end{figure}
            \begin{figure}
            	\centering
            	\begin{minipage}{.45\linewidth}
                    \centering
                    \includegraphics[width=\linewidth]{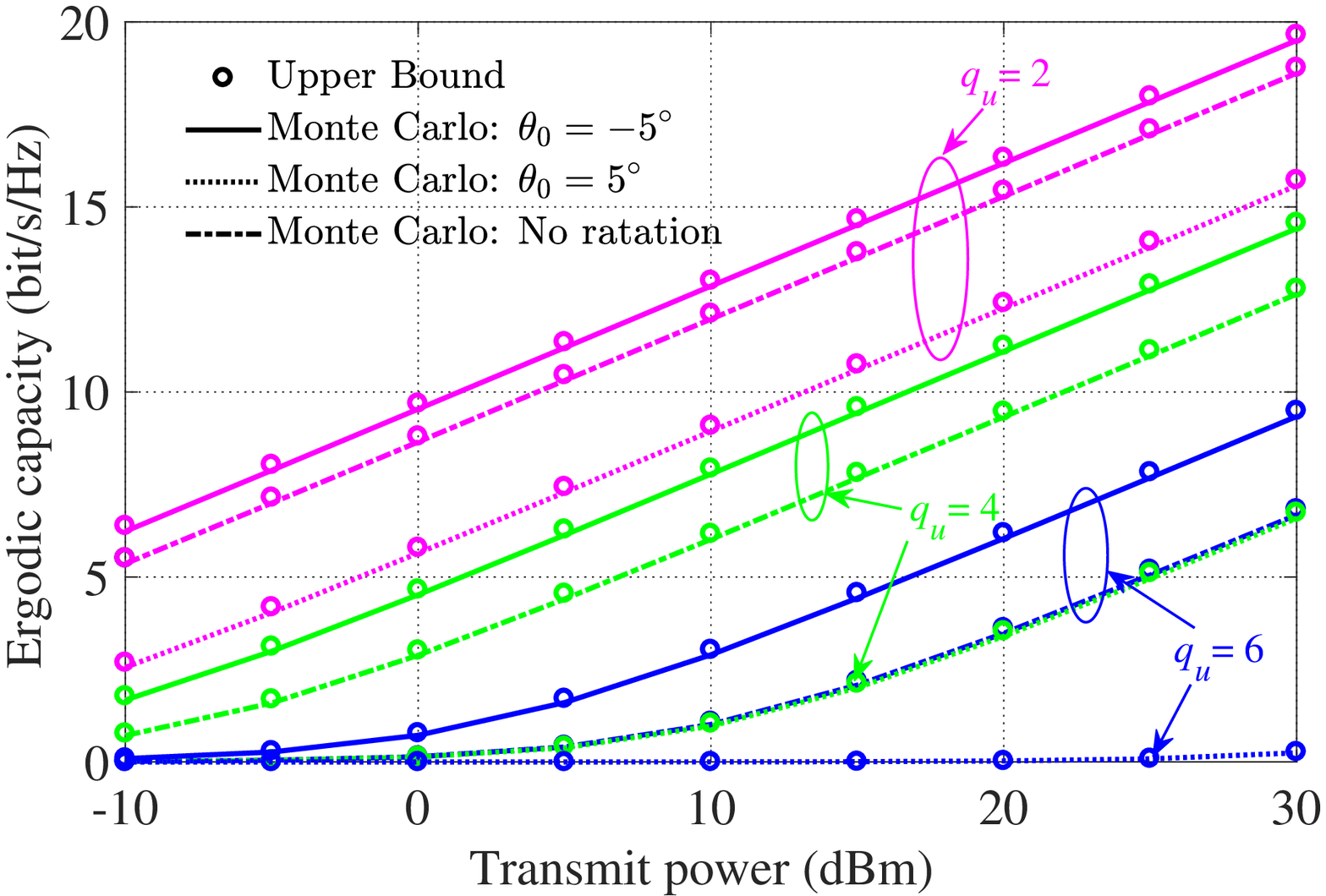}
                    \caption{Impact of $q_u$ on the ergodic capacity.}
                    \label{Fig_7}
            	\end{minipage}
                \hspace{0.2in}
            	\begin{minipage}{.45\linewidth}
                    \centering
                    \includegraphics[width=\linewidth]{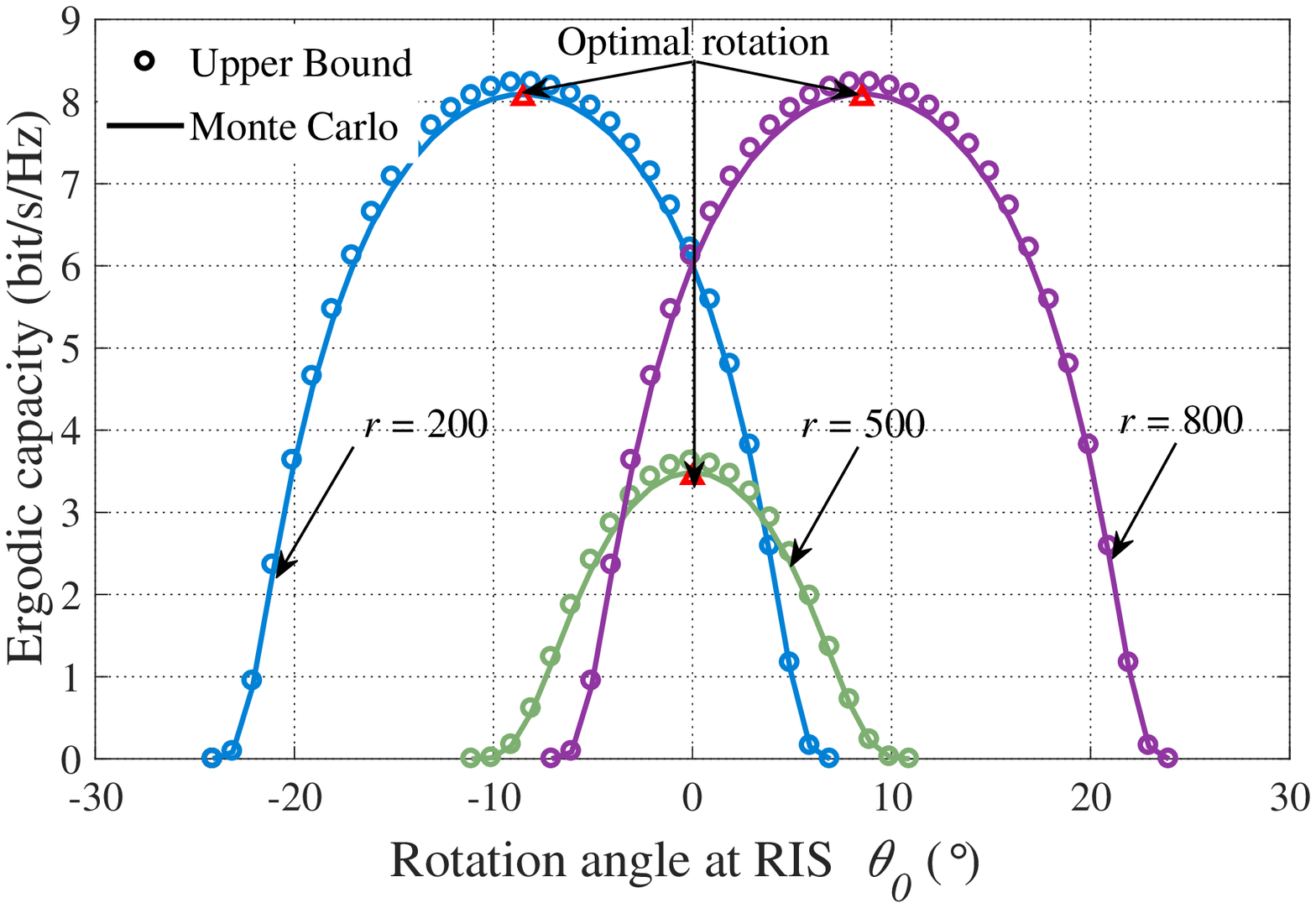}
                    \caption{Impact of $\theta_{0}$ on the ergodic capacity.}
                    \label{Fig_8}
            	\end{minipage}
            \hspace{0.2in}
            	\begin{minipage}{.45\linewidth}
                    \centering
                    \includegraphics[width=\linewidth]{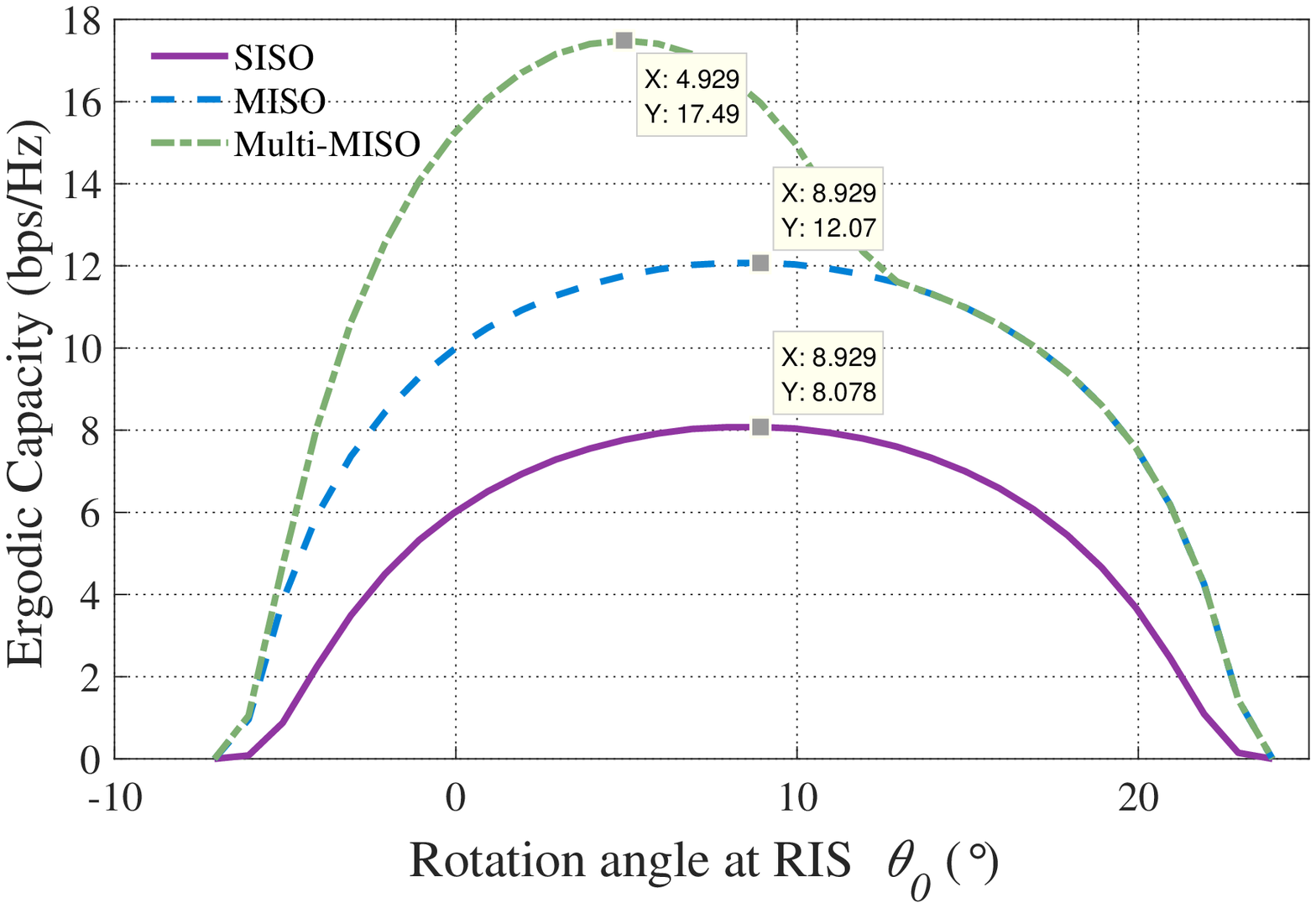}
                    \caption{Optimal $\theta_{0}$ in different scenarios.}
                    \label{Fig_9}
            	\end{minipage}
            \end{figure}
            Fig. \ref{Fig_7} shows the impact of directivity parameter of RIS ($q_{u}$) on the ergodic capacity with different rotations at the RIS, including $\theta_{0}=5^{\circ}$ and $\theta_{0}=-5^{\circ}$. In general, a larger $q_{u}$  means a stronger directivity and higher gain. Hence, there will be a better performance by leveraging a RIS with a stronger directivity. However, as plotted in Fig. \ref{Fig_7}, the performance deteriorates sharply as $q_{u}$ increases. This may be caused by the narrower beamwidth when the directivity becomes stronger. Accordingly, it is difficult to obtain considerable gain in the direction of the Tx-Ant and Rx-Ant meanwhile as discussed in \emph{Remark \ref{remark:RIS-rotation-performance}}. As for the rotation at RIS, the performance with $5^{\circ}$ rotation counterclockwise, i.e. $\theta_{0}=-5^{\circ}$, is significantly improved than that without rotation, while the performance with $5^{\circ}$ rotation clockwise, i.e. $\theta_{0}=5^{\circ}$, gets worse. This also reveals that the correct and reasonable RIS rotation adjustment is of great importance to the performance.

            Moreover, Fig. \ref{Fig_8} presents the impact of the RIS rotation angle ($\theta_{0}$) on the ergodic capacity. In the figure, we illustrate three cases (locations), including $P(100,200,100)$, $P(100,500,100)$, and $P(100,800,100)$. As can be seen, the maximum capacities are obtained with the RIS rotation angles of $-8.5^{\circ}$, $0^{\circ}$, and $8.5^{\circ}$, respectively. Correspondingly, this indicates that the optimal rotation angles at the RIS are $-8.5^{\circ}$, $0^{\circ}$, and $8.5^{\circ}$, respectively. This is consistent with the theoretical analysis in \emph{Theorem \ref{theorem:cascade-channel-gain-rotation-max}}. According to (\ref{eq:varphi-TxRISc-AOA-RIScRx-AOD}), we can obtain $\theta_{t,c}^{\sss{\text{AOA}}}=65.9052^{\circ}, 78.9042^{\circ}, 82.9294^{\circ}$ and $\theta_{c,r}^{\sss{\text{AOD}}}=82.9294^{\circ}, 78.9042^{\circ},65.9052^{\circ}$, respectively. Hence, the optimal rotation angles are $\left(65.9052-82.9294\right)/2=-8.5121$, $\left(78.9042-78.9042\right)/2=0$ and $\left(82.9294-65.9052\right)/2=8.5121$, respectively. Besides, we observe that the capacity becomes worse with the other rotation angles and even tends to zero when the angle exceeds a certain range. This may result from the obstructing of the signal caused by improper RIS rotation, which verifies the conclusions in \emph{Property \ref{property:RIS-rotation-constrain}}.

            As depicted in Fig. \ref{Fig_9}, we compare the impact of the RIS rotation angle on the capacity in different scenarios, including SISO, MISO and Multi-MISO scenarios. There are two users considered with the location $(100,1000,100)$ and $(100,1200,100)$ in the multi-MISO scenario, while only the user at $(100,1000,100)$ in considered in SISO and MISO scenarios. In addition, the RIS is assumed located at $(0,800,100)$. As can be seen, in SISO and MISO scenarios, the optimal rotation angles of the RIS are both about $8.926^\circ$, which is consent with the theoretical analysis in \emph{Theorem \ref{theorem:cascade-channel-gain-rotation-max}}. This is because, even considering multiple antennas those are located in the far-field region of RIS, the AOAs from the antennas to the RIS are nearly equal. Hence, for the system with multiple antennas, the conclusions are still applicable. As for the scenario with two users, the optimal rotation angle is different and it is smaller than those in the other two scenarios. This indicates that the RIS is oriented to the middle of the users to ensure the performance of each user. Please note that when the rotation angle exceeds a certain value, the ergodic capacity coincides with that of the MISO scenario. This phenomenon is caused by the fact that one of the user is in the back of the RIS when the RIS is rotated with a larger angle.
        %    \begin{figure}[!t]
    %            \centering
    %            \includegraphics[width=0.85\linewidth]{./Figs/TxRxRIS_Rotation.eps}
    %            \caption{Impact of RIS rotation on the ergodic capacity.}
    %            \label{fig:Capacity-rotation-ris}
    %        \end{figure}
      %      \begin{figure}[!t]
    %            \centering
    %            \includegraphics[width=0.85\linewidth]{./Figs/Optimal_rotation_ris_r.eps}
    %            \caption{Optimal rotation at the RIS.}
    %            \label{fig:Optimal_rotation_ris}
    %        \end{figure}

            \begin{figure}
            	\centering
            	\begin{minipage}{.45\linewidth}
                    \centering
                    \includegraphics[width=\linewidth]{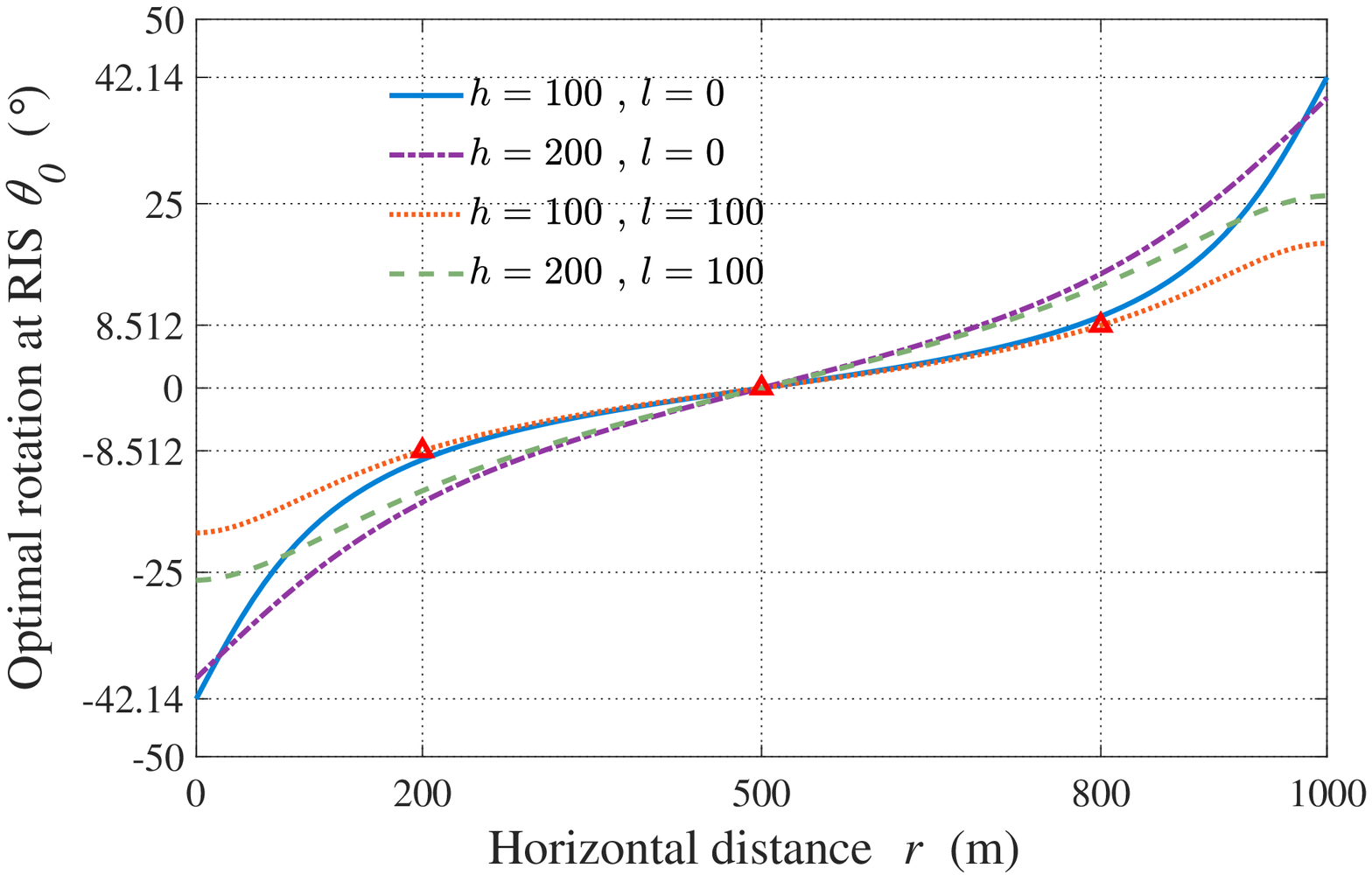}
                    \caption{Optimal rotation angle at the RIS.}
                    \label{Fig_10}
            	\end{minipage}
                \hspace{0.2in}
            	\begin{minipage}{.45\linewidth}
                    \centering
                    \includegraphics[width=\linewidth]{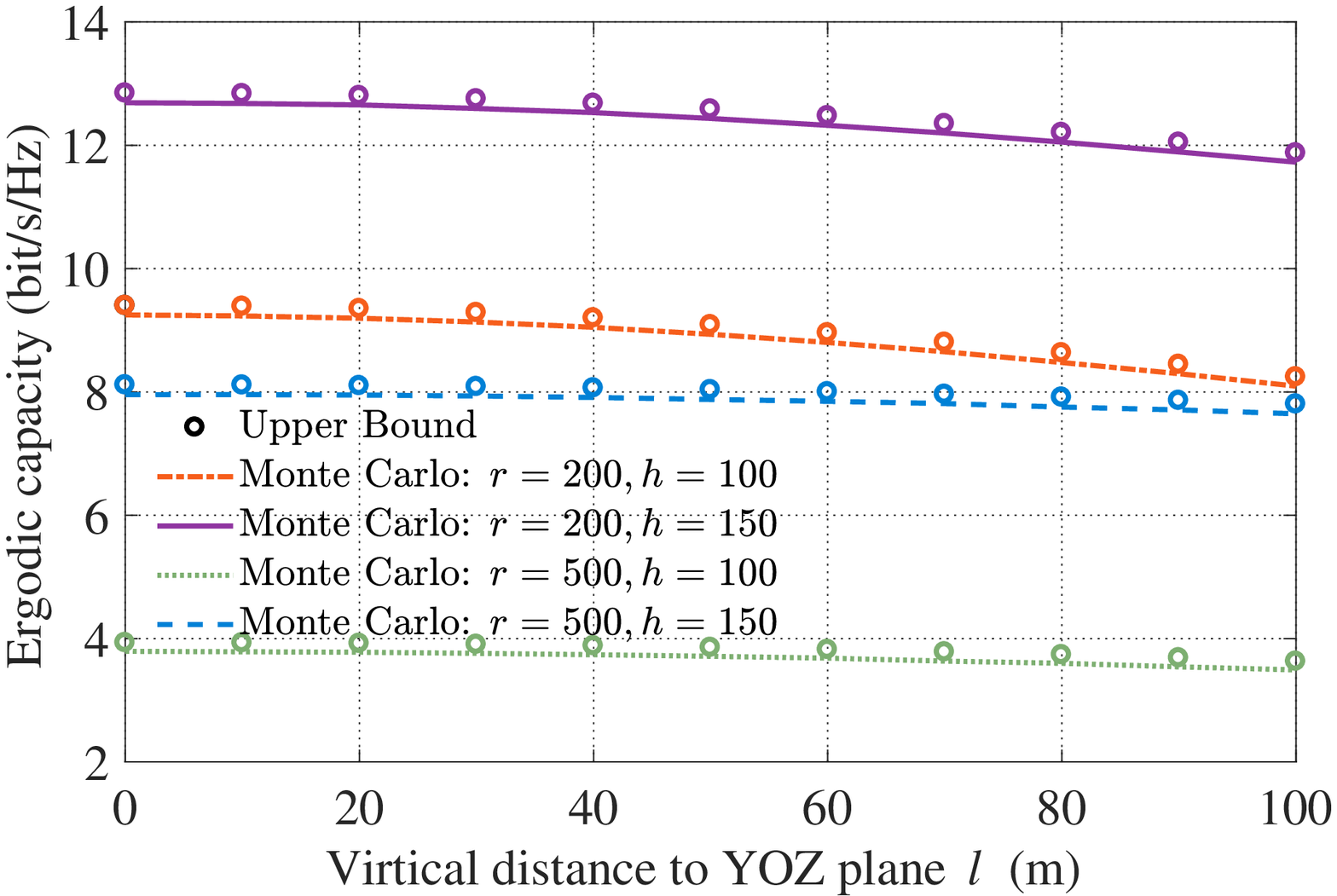}
                    \caption{Impact of $l$ on the ergodic capacity.}
                    \label{Fig_11}
            	\end{minipage}
            \end{figure}
            Fig. \ref{Fig_10} depicts the optimal rotation angle at the RIS as concluded in (\ref{eq:ris-rotation-opt}) and (\ref{eq:ris-rotation-opt-location-dep}). Based on the previous introduction, it is clear that a smaller $r$ represents a smaller horizontal distance from the RIS to the Tx-Ant, i.e. the RIS is closer to the Tx-Ant than to the Rx-Ant. As can be seen from the figure, the RIS ought to be rotated counterclockwise by a certain angle, i.e. $\theta_{0}<0$, when the RIS is closer to the Tx-Ant, while to be rotated clockwise by an angle, i.e. $\theta_{0}>0$, when the RIS is closer to the Rx-Ant. For example, the optimal rotation angles are $-8.512^{\circ}$ and $8.512^{\circ}$, respectively, at the two different locations $P(100,200,100)$ and $P(100,800,100)$.  Such conclusion is consistent with the illustration in Fig. \ref{Fig_8}. Additionally, the optimal rotation is $0^{\circ}$ when the RIS locates above the middle of the Tx-Rx pair, i.e. $r=500$ m. This means that, for any altitude, when the RIS is above the middle of the Tx-Rx pair, the RIS should be kept parallel to the XOY plane, as described in \emph{Remark \ref{remark:RIS-rotation-opt}}.

        \subsection{Impact of the RIS location on the capacity}
            Fig. \ref{Fig_11} shows the impact of the vertical distance from the RIS to the YOZ plane, i.e. $l$, on the ergodic capacity. We investigate the trend of the capacities by increasing $l$ in four cases, including ($r=200,h=100$), ($r=200,h=150$), ($r=500,h=100$), and ($r=500,h=150)$. As illustrated, an increasing $l$ results in a performance degradation indicating that the RIS should be deployed directly above the Tx-Rx pair, i.e. $l=0$, as analyzed in \emph{ Corollary \ref{corollary:l-opt}}. Hence, $l=0$ will be adopted in the following simulations unless otherwise stated.
    %    \begin{figure}[!t]
%            \centering
%            \includegraphics[width=0.85\linewidth]{./Figs/ErgodiCapacity_L.eps}
%            \caption{Impact of $l$ on the ergodic capacity.}
%            \label{fig:Capacity―L}
%        \end{figure}

\begin{figure}
                	\centering
                	\begin{minipage}{.45\linewidth}
                        \centering
                        \includegraphics[width=\linewidth]{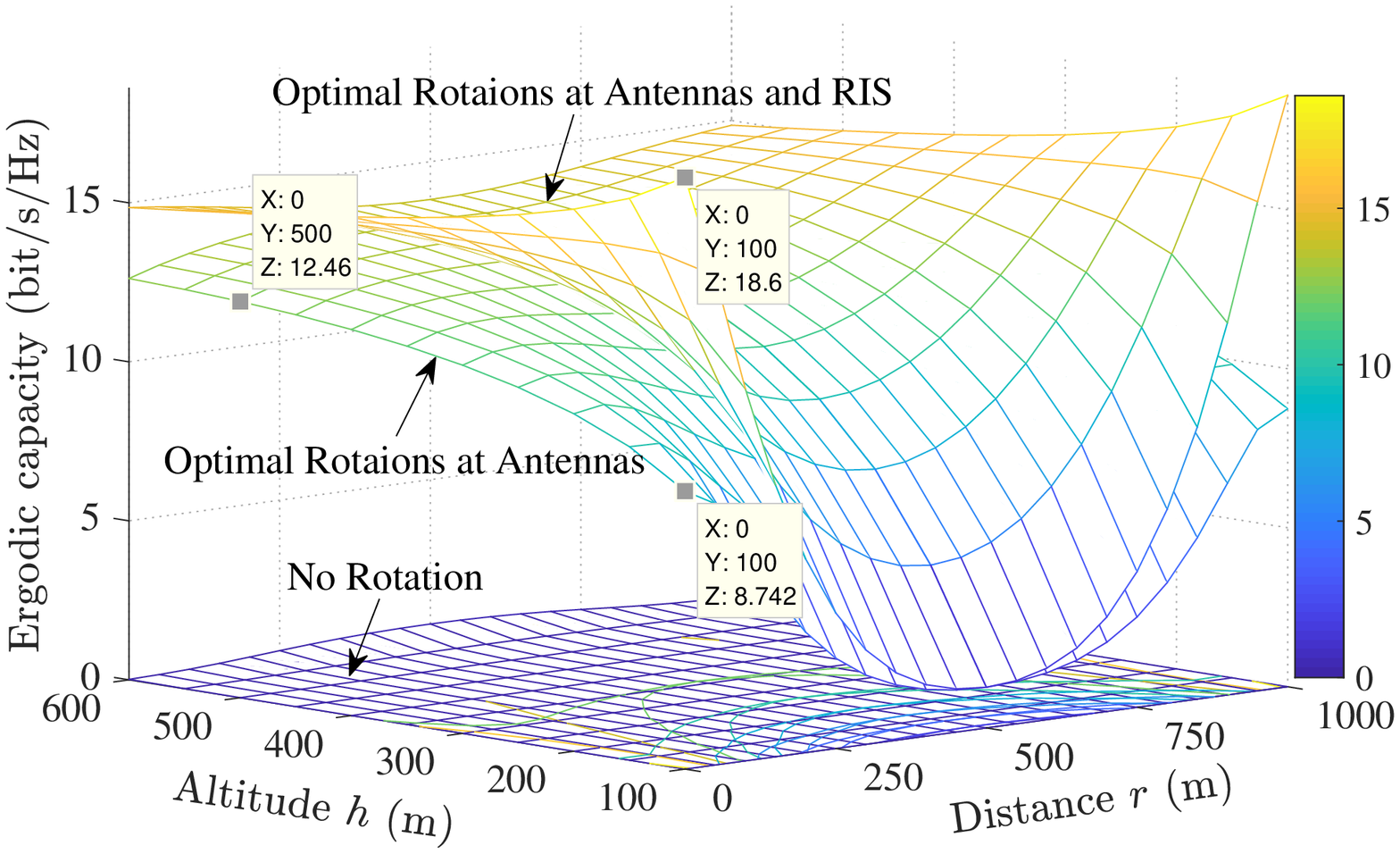}
                        \caption{Impact of $h$ and $r$ on the ergodic capacity.}
                        \label{Fig_12}
                	\end{minipage}
                    \hspace{0.2in}
                	\begin{minipage}{.45\linewidth}
                    \centering
                    \includegraphics[width=\linewidth]{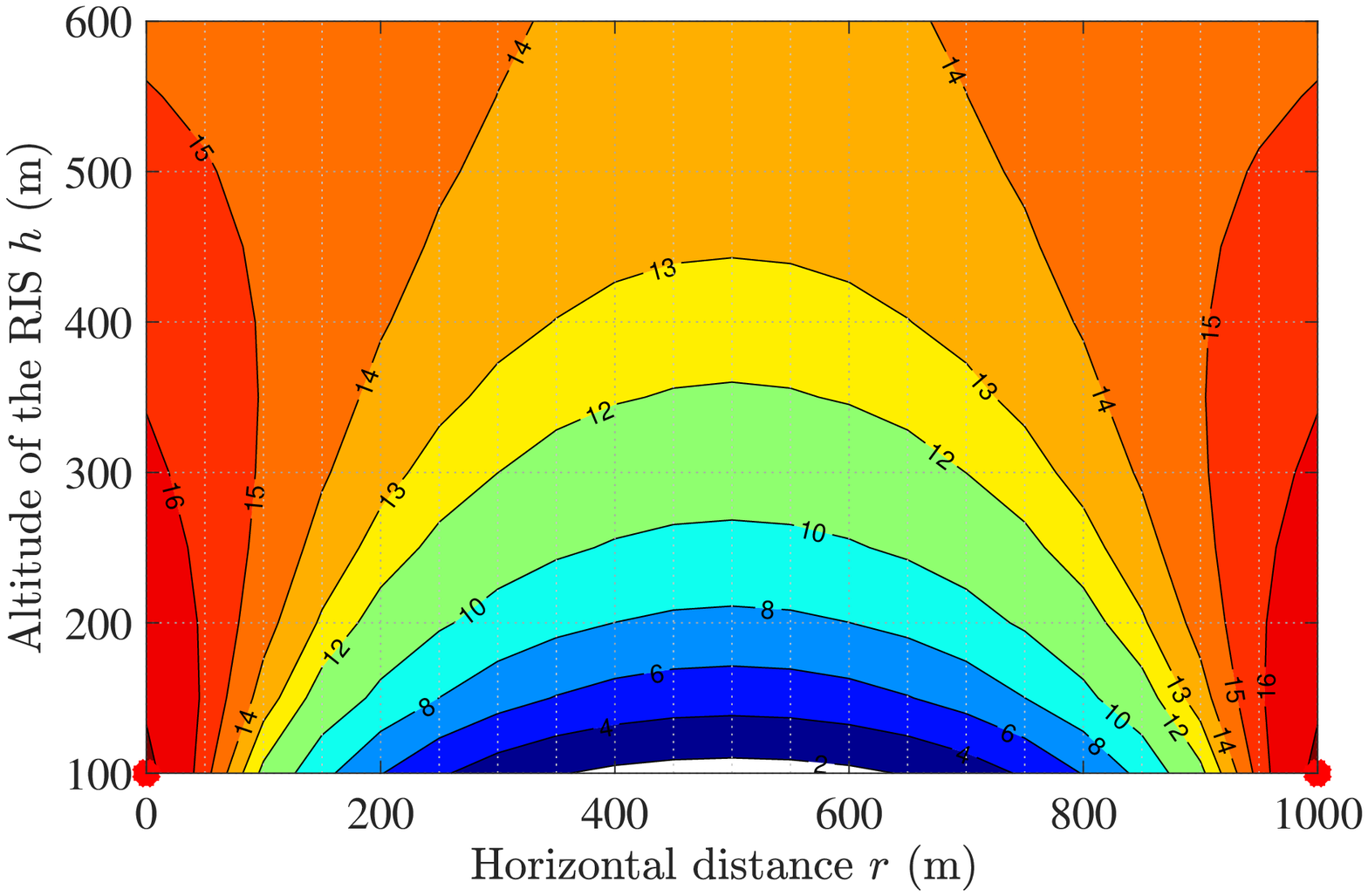}
                    \caption{Optimal location and effective region.}
                    \label{Fig_13}
                	\end{minipage}
                \end{figure}

            Furthermore, Fig. \ref{Fig_12} compares the ergodic capacities of different locations under three different conditions: (1) no rotation at the antennas and RIS: there is no rotation at the antennas (both Tx-Ant and Rx-Ant) and the RIS with their initial directions in the positive and negative of the Z-axis, respectively; (2) the optimal rotations at the antennas: both the Tx-Ant and Rx-Ant are with optimal rotations pointing to the RIS center, while RIS is in the initial direction without rotation; (3) the optimal rotations at the antennas and RIS: the Tx-Ant and Rx-Ant are with optimal rotations, and the RIS rotates with the optimal angle as (\ref{eq:ris-rotation-opt}). As we can see, when the antennas and RIS rotate with optimal angles, the performance surpasses those in the other two cases. To be specific, there is a significant improvement with optimal rotations at the antennas comparing to the case without any rotations, in which the communications may be interrupted as the capacity tends to zero. Additionally, there is also a big gap between the performance with only optimal rotation at the antennas and that with optimal rotations at the antennas and RIS. For instance, at $P(0,0,100)$, the capacity is less than 10 bit/s/Hz with only optimal rotation at the antennas, whilst it exceeds 15 bit/s/Hz with optimal rotation at the antennas and RIS. This proves the necessity of appropriately adjusting the antennas and RIS rotations in guaranteeing a high performance. Please note that there is the same performance when $r=500$ in the last two conditions. This is because the theoretical optimal rotation $\theta_{0}$ equals $0$ when $r=R/2$. This again verifies the correctness of \emph{Remark \ref{remark:RIS-rotation-opt}}, and there is no need to rotate the RIS when the RIS is exactly located above the middle of the Tx-Rx pair.
       %     \begin{figure}[!t]
    %            \centering
    %            \includegraphics[width=0.85\linewidth]{./Figs/ErgodiCapacity_Location.eps}
    %            \caption{Ergodic capacity.}
    %            \label{fig:Capacity-lacatoin}
    %        \end{figure}

            We also notice that the RIS location have a significant impact on the performance even with optimal rotations. Concretely, for a fixed $r$, e.g. $r=250$, a smaller $h$ will result in a worse performance. This seems to be counterintuitive because a smaller $h$ may lead to a shorter distance, thereby resulting in less path loss and better performance in turn. Nevertheless, when the RIS is closer to the XOY plane, i.e. a smaller $h$, the degree of parallelism between the RIS and XOY plane gets higher. As a result, the Tx-Ant and Rx-Ant may be located at the side lobe (poor directon) of the RIS meantime. Naturally, there is no doubt that the performance will deteriorate. As for the $r$, an increasing $r$ incurs an eroded performance when $r\in\left[0,R/2\right]$, while the effect is opposite when $r\in\left[R/2,R\right]$. This suggests that we can achieve the maximum performance when $r=0$ or $r=R$. Though this effect is evident when $h$ is small, it becomes less obvious or even changes as $h$ increases. This phenomenon is caused by the relative magnitude of $h$ and $R$ as discussed previously in \emph{Corollary \ref{corollary:r-opt}}. Additionally, it is obvious that the rotation of the RIS is of great significance to improve the performance, especially when the RIS is close to the Tx-Ant or Rx-Ant. For example, we can achieve more than 200\% performance improvement (from 8.742 to 18.60 bit/s/Hz) by rotating the RIS $42.14^{\circ}$ (when $r=0$ as depicted in the blue solid curve in Fig. \ref{Fig_10}) counterclockwise at $P(0,0,100)$, whilst less than 150\% improvement (from 8.742 to 12.46 bit/s/Hz) by moving the RIS over 400 meters vertically.

            Finally, Fig. \ref{Fig_13} presents the deployment of the RIS. As stated at the beginning of this section, the RIS can be moved from the Tx-Ant to Rx-Ant, i.e. $r\in[0,R]$, at the altitude $h\in[h_{\text{min}},h_{\text{max}}]$. Please note that for the optimization of the RIS location and effective region, the capacity of each location is based on the optimal rotation as given in (\ref{eq:ris-rotation-opt-location-dep}). Based on \emph{Algorithm 1}, we obtain the best two locations, including $P(0,0,h_{\text{min}})$ and $P(0,R,h_{\text{min}})$, for maximum capacity. To illustrate the effectiveness of \emph{Algorithm 2}, the SNR threshold is set as $\gamma_{\text{th}}=45$ dB corresponding the capacity of 15 bit/s/Hz. As can be seen, the RIS should be deployed at the red area to guarantee the given SNR at the Rx-Ant. Last but not the least, we observe that increasing the altitude $h$ has less impact on the performance than changing the horizontal distance $r$. Combining the discussion in Fig. \ref{Fig_12}, we can obtain a considerable performance at a low altitude when the RIS is deployed closely to the Tx-Ant or Rx-Ant with the corresponding optimal rotations.
  %      \begin{figure}[!t]
%            \centering
%            \includegraphics[width=0.85\linewidth]{./Figs/Deployement.eps}
%            \caption{Optimal location and suitable area.}
%            \label{fig:Deployement}
%        \end{figure}

    \section{Conclusion}
        In this paper, we provided a new degree of freedom in optimizing the RIS-aided wireless communication channel by studying the radiation characteristics of RIS. We derived a tight upper bound of the ergodic capacity over cascade Rician fading channels, which applies to a RIS of any scale, and found that the orientations of antennas and RIS play a vital role in the performance. On this basis, we extracted the optimal rotations for any location of the RIS. Furthermore, we deduced the location-dependent expressions and investigated the optimal locations and effective regions for the RIS deployment. Most importantly, we verified that it is more efficient to rotate the RIS than move it over a wide area. We believe that the presented analyses in this paper can provide insightful references and guidelines for the design and deployment of RIS-aided wireless communication systems in the future.
    \appendices
        \section{}\label{proof:capacity-upperbound-theory}
        We firstly provide some mathematical results about a random variable $X=\left|a u+b v\right|$, where $u$ is a complex constant; $v$ is a random variable with $v\sim \mathcal{CN}(0,1)$; $a>0$ and $b>0$. By defining $Z=b^{-2}\left(a u+b v\right)^{2}=b^{-2}X^2$, $2Z$ fellows the non-central chi-squared distribution with 2 degrees of freedom, i.e. $2Z\sim \chi_{2}^{2}\left(2\lambda\right)$, where $\lambda=a^{2}b^{-2}\left|u\right|^{2}$. Therefore, we can obtain the possibility distribution function of $Z$ as \cite[Eq. (406)]{ALapidothCapacitybounds}
         \begin{equation}
            f_{Z}\left(z\right)=e^{-z-\lambda}I_{0}\left(2\sqrt{\lambda z}\right),
         \end{equation}
        with $\mathbb{E}(Z)=1+\lambda$, where $I_{0}(\cdot)$ is the zero-th order modified Bessel functions of the first kind with $I_{\nu}(x)=\sum_{j=0}^{\infty} \frac{1}{j!\Gamma(\nu+j+1)}\left(\frac{x}{2}\right)^{\nu+2j}$\cite[Eq. (8.445)]{ISGTableofIntegrals}.

        Further, the possibility distribution function and expectation of $X$ are respectively given by
            \begin{equation}\label{eq:Nocentr-combi-sqrt-Chi-square-Comp-distr-PDF}
                \begin{aligned}
                    f_{X}(x)&=\frac{2}{b^{2}}xf_{Z}(\frac{1}{b^{2}}x^2)
                    =\frac{2}{b^{2}}x e^{-{\frac{x^{2}}{b^{2}}}-\lambda}I_{0}\left(\frac{2\sqrt{\lambda}}{b}x\right),
                \end{aligned}
            \end{equation}
            \begin{equation}\label{eq:Nocentr-combi-sqrt-Chi-square-Comp-distr-Exp-proof}
            \begin{aligned}
            \mathbb{E}(X)&=\int_{0}^{\infty}\frac{2}{b^{2}}x^{2} e^{-{\frac{x^{2}}{b^{2}}}-\lambda}I_{0}\left(\frac{2\sqrt{\lambda}}{b}x\right)d_{x}
                \overset{\mathcal{A}}{=}e^{-\lambda}\frac{2}{b^{2}}\sum_{j=0}^{\infty} \frac{\lambda^{j}}{j!\Gamma(j+1)\beta^{2j}}\int_{0}^{\infty}x^{2+2j}e^{-\frac{1}{b^{2}}x^{2}}d_{x}\\
                &\overset{\mathcal{B}}{=}e^{-\lambda}\frac{2}{b^{2}}\sum_{j=0}^{\infty} \frac{\lambda^{j}}{j!\Gamma(j+1)\beta^{2j}}\frac{b^{{3+2j}}}{2}\Gamma(j+\frac{3}{2})
                \overset{\mathcal{C}}{=}e^{-\lambda}b\frac{\Gamma(\frac{3}{2})}{\Gamma(1)}{_{1}F_{1}}\left(\frac{3}{2};1;\lambda\right),
            \end{aligned}
            \end{equation}
            where process $\mathcal{A}$ is obtained by expressing $I_{0}\left(2b^{-1}\sqrt{\lambda}x\right)$ as a power series \cite[Eq. (8.445)]{ISGTableofIntegrals}. Moreover, $\mathcal{B}$ and $\mathcal{C}$  follow \cite[Eq. (3.326-2)]{ISGTableofIntegrals} and \cite[Eq. (15.1.1)]{MAbramowitzHandbook}, respectively. Therefore, leveraging \cite[Eq. (8.331-1), Eq. (8.338-2)]{ISGTableofIntegrals}, the expectations of $X$ and  $X^{2}$ are respectively obtained as
           \begin{equation}\label{eq:Nocentr-combi-sqrt-Chi-square-Comp-distr-Exp}
            \mathbb{E}\left\{X\right\}=\frac{b\sqrt{\pi}e^{-\lambda}}{2}{_{1}F_{1}}\left(\frac{3}{2};1;\lambda\right),
           \end{equation}
           \begin{equation}\label{eq:Nocentr-combi-Chi-square-Comp-distr-Exp}
            \mathbb{E}\left\{X^{2}\right\}=b^{2}\left(\lambda+1\right).
           \end{equation}

        Based on Jensen's inequality, we have
            \begin{equation} \label{eq:capacity-downlink-upbound-proof}
                \begin{aligned}
                 C\leq\log_{2}\left(1+\mathbb{E}\left\{\mathcal{SNR}\right\}\right).
                \end{aligned}
            \end{equation}
            As for $\mathbb{E}\left\{\mathcal{SNR}\right\}$, we have the following conclusions according to (\ref{eq:Nocentr-combi-sqrt-Chi-square-Comp-distr-Exp}) and (\ref{eq:Nocentr-combi-Chi-square-Comp-distr-Exp}),
            \begin{equation} \label{eq:hn-norm-Exp}
                \begin{aligned}
                    \mathbb{E}\left\{\left|g_{n}\right|\right\}&=\sqrt{\frac{\pi\rho_{t,n}}{1+K_{1}}}\frac{{_{1}F_{1}}\left(\frac{3}{2};1;K_{1}\right)}{2 e^{K_{1}}},n\in\left(1,\dots,N\right),
                \end{aligned}
            \end{equation}
            \begin{equation} \label{eq:zn-norm-Exp}
                \begin{aligned}
                    \mathbb{E}\left\{\left|z_{n}\right|\right\}&=\sqrt{\frac{\pi\rho_{n,r}}{1+K_{2}}}\frac{{_{1}F_{1}}\left(\frac{3}{2};1;K_{2}\right)}{2 e^{K_{2}}},n\in\left(1,\dots,N\right),
                \end{aligned}
            \end{equation}
            \begin{equation} \label{eq:hn-square-Exp}
                \begin{aligned}
                    \mathbb{E}\left\{\left|g_{n}\right|^{2}\right\}&=\rho_{t,n},n\in\left(1,\dots,N\right),
                \end{aligned}
            \end{equation}
            \begin{equation} \label{eq:zn-square-Exp}
                \begin{aligned}
                    \mathbb{E}\left\{\left|z_{n}\right|^{2}\right\}&=\rho_{n,r},n\in\left(1,\dots,N\right).
                \end{aligned}
            \end{equation}
            Additionally, we assume that there is no power loss of the reflection at the RIS, i.e. $\left|\mathcal{T}\right|=1$ \cite{YGaoDistrRISMISO,SZhangRISCapacityDeployment}. So, the expression of $\mathbb{E}\left\{\mathcal{SNR}\right\}$ can be described as
            \begin{equation} \label{eq:SNR-max-Exp}
                \begin{aligned}
                    \mathbb{E}\left\{\mathcal{SNR}\right\}&=\frac{P_{\text{t}}}{N_{0}}\sum_{n=1}^{N}\mathbb{E}\left\{\left|g_{n}\right|^{2}\left|z_{n}\right|^{2}\right\}+\frac{P_{\text{t}}}{N_{0}}\sum_{n=1}^{N}\sum_{\substack{i=1\\i\neq n}}^{N}\mathbb{E}\left\{\left|g_{n}\right|\left|z_{n}\right|\left|g_{i}\right|\left|z_{i}\right|\right\}\\
                    &=\frac{P_{\text{t}}}{N_{0}}\sum_{n=1}^{N}\rho_{t,n}\rho_{n,r}+\frac{P_{\text{t}}}{N_{0}}\sum_{n=1}^{N}\sum_{\substack{i=1\\i\neq n}}^{N}\frac{\sqrt{\rho_{t,n}\rho_{t,i}\rho_{n,r}\rho_{i,r}}\pi^{2}{_{1}F_{1}}^{2}\left(\frac{3}{2};1;K_{1}\right){_{1}F_{1}}^{2}\left(\frac{3}{2};1;K_{2}\right)}{16\left(1+K_{1}\right)\left(1+K_{2}\right)e^{2\left(K_{1}+K_{2}\right)}}
                \end{aligned}
            \end{equation}

            In the far-field of the RIS, the distances between the Tx-Ant and each unit are usually supposed to be equal, i.e. $d_{t,n}=d_{t,c}$, with $d_{t,c}$ denoting the distance between the Tx-Ant and the RIS center. Analogously, we have $d_{n,r}=d_{c,r}$ with $d_{c,r}$ being the distance between the RIS center and the Rx-Ant. Definitely, we can get $\rho_{t,n}=\rho_{t,c}$, $\rho_{n,r}=\rho_{c,r}$ for $n\in\left(1,\dots,N\right)$ in the far-field of RIS. Also, we have $\rho_{t,n}\rho_{t,i}=\rho_{t,c}^{2}$ and $\rho_{t,n}\rho_{n,r}=\rho_{t,c}\rho_{c,r}$ for $i,n\in\left(1,\dots,N\right)$. Then, (\ref{eq:SNR-max-Exp}) can be simplified and the upper bound of (\ref{eq:capacity-downlink-upbound-proof}) is obtained. Then, the proof ends.

        \section{}\label{proof:optimal-roration-theory}
        According to (\ref{eq:cascade-channel-tx-rx-max}) and substituting $\alpha=\theta_{t,c}^{\sss{\text{AOA}}}$ and $\beta=\theta_{c,r}^{\sss{\text{AOD}}}$, the CCG with the RIS rotation can be rewritten as
                \newcounter{TempEqCnt11}
                \setcounter{TempEqCnt11}{\value{equation}}
                \setcounter{equation}{48}
        \begin{equation} \label{eq:cascade-channel-gain-rotation}
            \begin{aligned}
            \rho_{\text{cc}}^{\text{R}}&=\frac{\rho_{0}^{2}10^{0.2\left(q_{t}+q_{r}+2q_{u}+4\right)}}{d_{t,c}^{2}d_{c,r}^{2}}
                                      \cos^{q_{u}}\left(\alpha-\theta_{0}\right)\cos^{q_{u}}\left(\beta+\theta_{0}\right).
            \end{aligned}
        \end{equation}
        Let us discuss the function $f\left(x\right)=\cos\left(c-x\right)^{p}\cos\left(d+x\right)^{p}$, with $c\in\left(0,\pi/2\right); d\in\left(0,\pi/2\right)$; $p\geq1$ and $c-\pi/2<x<\pi/2-d$. The derivation of $f\left(x\right)$ is given by
        \begin{equation} \label{eq:derivation-func_f}
            f^{'}\left(x\right)=p\cos^{p-1}\left(c-x\right)\cos^{p-1}\left(d+x\right)\sin\left(c-d-2x\right).
        \end{equation}
        It is clear that the first three terms are all greater than zero. So, we resort to the property of the last term.

        When $c<d$, we have $c-\pi/2<(c-d)/2<0<\pi/2-d$. And we notice that $\sin\left(c-d-2x\right)>0$ when $x\in\left(c-\pi/2,(c-d)/2\right)$, and $\sin\left(c-d-2x\right)<0$ when $x\in\left((c-d)/2,\pi/2-d\right)$. This means that the maximum of $f\left(x\right)$ is obtained at $x=(c-d)/2$ when $c<d$. A similar proof procedure when $c>d$, and is omitted for brevity. The maximum of $f\left(x\right)$ is obtained when $x=(c-d)/2$, thus we have
        \begin{equation} \label{eq:product-cos-max}
            f_{\text{max}}=f\left(\frac{c-d}{2}\right)=\frac{\left[1+\cos\left(c+d\right)\right]^{p}}{2^{p}}.
        \end{equation}

        Hence, combining \emph{Property \ref{property:alpha-beta-range}}, \emph{Property \ref{property:RIS-rotation-constrain}} and (\ref{eq:product-cos-max}), the maximum CCG in (\ref{eq:cascade-channel-gain-rotation}) can be obtained when $\theta_{0}=(\alpha-\beta)/2$. Thus, the proof ends.

        \section{}\label{proof:r-optimal}

            By defining $\mu\left(r,h\right)=\left[1+\frac{r^{2}-Rr+h^{2}}{\sqrt{\left(r^{2}+h^{2}\right)\left[\left(R-r\right)^{2}+h^{2}\right]}}\right]^{q_{u}}$ and $\nu\left(r,h\right)=\frac{1}{\left(r^{2}+h^{2}\right)\left[\left(R-r\right)^{2}+h^{2}\right]}$, (\ref{eq:cascade-channel-gain-rotation-dist}) can be rewritten as
            \begin{equation} \label{eq:cascade-channel-gain-rotation-dist-proof}
            \begin{aligned}
                \rho_{\text{cc}}^{\text{R}}\left(r,h\right)&=\frac{\rho_{0}^{2}10^{0.2\left(q_{t}+q_{r}+2q_{u}+4\right)}}{2^{q_{u}}}\mu\left(r,h\right)\nu\left(r,h\right).
            \end{aligned}
            \end{equation}
            Furthermore, we have the derivations as following
            \begin{equation} \label{eq:gx-derivation}
            \begin{aligned}
                \frac{\partial \mu}{\partial r}&=\frac{q_{u}h^{2}R^{2}\left(2r-R\right)}{\left(r^{2}+h^{2}\right)^{\frac{3}{2}}\left[\left(R-r\right)^{2}+h^{2}\right]^{\frac{3}{2}}}
                \left[1+\frac{r^{2}-Rr+h^{2}}{\sqrt{\left(r^{2}+h^{2}\right)\left[\left(R-r\right)^{2}+h^{2}\right]}}\right]^{q_{u}-1},
            \end{aligned}
            \end{equation}
            \begin{equation} \label{eq:gx-derivation}
            \begin{aligned}
                \frac{\partial \nu}{\partial r}=\frac{2\left(R-2r\right)\left(r^{2}-Rr+h^{2}\right)}{\left(r^{2}+h^{2}\right)^{2}\left[\left(R-r\right)^{2}+h^{2}\right]^{2}}.
            \end{aligned}
            \end{equation}
            Based on \emph{Remark \ref{remark:cos-alpha-plus-beta}}, we notice that $\partial \mu/\partial r<0$ when $r\in\left(0,R/2\right)$, and $\partial \mu/\partial r>0$ when $r\in\left(R/2,R\right)$. So, for a fixed $h$, $\mu\left(h,r\right)$ reaches its maximum at $r=0$ or $r=R$, and its minimum at $r=R/2$. As for $\nu\left(r,h\right)$, we firstly discuss the case of $R>2h$. According to (\ref{eq:gx-derivation}), we have
            $\partial \nu/\partial r>0$ when $r\in\left[0,r_{1}\right]\cup\left[R/2,r_{2}\right]$, and $\partial \nu/\partial r<0$ when $r\in\left[r_{1},R/2\right]\cup\left[r_{2},R\right]$, where $r_{1}=(R-\sqrt{R^{2}-4h^{2}})/2$ and $r_{2}=(R+\sqrt{R^{2}-4h^{2}})/2$.

            Therefore, in the case of $R>2h$, for a fixed $h$, both $\mu\left(r,h\right)$ and $\nu\left(r,h\right)$ are monotonically decreasing when $r\in\left[r_{1},R/2\right]$, and monotonically increasing when $r\in\left[R/2, r_{2}\right]$. Hence, when $r\in\left[r_{1},r_{2}\right]$, the local maximum of $\rho_{\text{cc}}^{\text{R}}\left(r,h\right)$ is obtained at $r=r_{1}$ or $r=r_{2}$. However, when $r\in\left[0, r_{1}\right]$ and $r\in\left[r_{2},R\right]$, the optimal $r$ for maximizing $\rho_{cc}\left(r,h\right)$ can not be determined because in these two intervals $\mu\left(r,h\right)$ is monotonically decreasing while $\nu\left(r,h\right)$ is monotonically increasing.

            Building on the above insights into the monotonicity of $\mu\left(r,h\right)$ and $\nu\left(r,h\right)$,  and their symmetry about $r=R/2$, the maximum exists in these two intervals, i.e. $r\in\left[0, r_{1}\right]$ and $r\in\left[r_{2},R\right]$. Hence, the conclusion in \emph{Case \ref{case:R-largethan-sqrtl2h2}} is proofed. In particular, when $R\gg2h$, we have $r_{1}\approx0$ and $r_{2}\approx R$. Thus, the optimal location is obtained as described in \emph{Case \ref{case:R-muchlargethan-sqrtl2h2}}.

            On the contrary, when $R< 2h$, $\nu\left(r,h\right)$ is monotonous increasing and decreasing when $r\in\left[0,R/2\right]$ and $r\in\left[R/2, R\right]$, respectively. The  monotonicity of $\mu\left(r,h\right)$ is reversed in these two intervals. Hence, it is difficult to figure out the optimal value $r$ directly as stated in \emph{Case \ref{case:R-lessthan-sqrtl2h2}}.

 %   \footnotesize % 调了顺序后，参考文献字体变大，回归正常字体
    %\bibliographystyle{IEEEtran}
%    \bibliography{IEEEabrv,mybibfile}
%\clearpage

   % \bibliographystyle{IEEEtran}
%    \bibliography{refs_yajun}


%\end{IEEEbiography}
%% insert where needed to balance the two columns on the last page with
%%\newpage


%\vfill



\end{document}